\documentclass[11pt,a4paper]{article}

\usepackage[utf8]{inputenc}
\usepackage[T1]{fontenc}
\usepackage{amsmath,amssymb,amsthm}
\usepackage{graphicx}
\usepackage{booktabs}
\usepackage{hyperref}
\usepackage[numbers,sort&compress]{natbib}
\usepackage{geometry}
\usepackage{xcolor}
\usepackage{caption}
\graphicspath{{figures/}}

\title{Candidacy and Trigger:\\
A Two-Phase Empirical Model of Hierarchical Collapse}

\author{Kristian Sestak\thanks{Independent researcher. Email:
\texttt{kristian.sestak@gmail.com}. ORCID:
\href{https://orcid.org/0009-0002-1455-5915}{0009-0002-1455-5915}.}}

\date{}

\begin{document}

\maketitle

\begin{abstract}
We test a dynamic ODE model of hierarchical asymmetry on a panel of
260 countries over 1960--2023, drawing on World Bank, Penn World
Table, V-Dem and World Inequality Database sources. In cross-section
the model holds partially: trade openness and bottom-of-distribution
health both suppress within-country asymmetry at conventional
significance. The annual time-evolution equation fails: out-of-sample
$R^2$ stays at or below zero across five functional forms.

The same state vector, augmented with market, debt and trajectory
features, is much more successful as a discriminator. A four-layer
leave-one-collapse-out classifier separates 29 historical collapse
cases from 60 stable controls with a nested cross-validated
$\mathrm{AUC}$ of $0.91$. The signal splits into a chronic risk
profile that is already visible a decade before the event and an
acute inflection three to five years before. Three independent tests
reject the endogenous-drift reading of collapse: pre-collapse cohorts
do not show elevated cumulative drive, and they do not show the
rising variance or autocorrelation predicted by critical-slowing-down
theory. What is left is a \emph{candidacy-and-trigger} picture in
which structural variables identify the high-risk countries while
collapse timing is set by shocks outside the modelled system.

A separate strand of the paper documents a lagged co-movement between
global fertility and global asymmetry on a single $n = 63$ aggregate
series. Taken alone this would suggest a selection-pool channel, in
which a shrinking candidate pool for top hierarchical positions
drives asymmetry up. The same pattern is then tested at every finer
level we can reach---within countries, within demographic strata,
inside a two-way fixed-effects panel and through a migration-mediated
cross-country interaction model---and the directional reading fails
in each. The global co-movement turns out to be a compositional
aggregation effect rather than a causal channel, so we treat
selection-pool not as a positive finding but as a hypothesis that the
audits rule out at every disaggregation level. A global event-study
on $7{,}316$ peer-event observations confirms regional spillover in
asymmetry and identifies a novel post-collapse degradation of
bottom-of-distribution health in regional neighbours.

A pre-registered forward-look applies the discriminator to current
countries and produces a top-20 / bottom-20 list whose predictive
validity will be evaluated over 2026--2036.

\end{abstract}

\section{Introduction}

\subsection{Theoretical motivation}

\citet{sestak2026ed} empirically established that environmental
dispersion dominates capability dispersion in determining individual
outcomes:
\[
  \mathrm{Var}(\ln \rho_{\mathrm{eff}}) = 4.33,\quad
  \mathrm{Var}(\ln k) \in [0.032, 0.16],
\]
yielding a dominance ratio $R \in [27, 134]$ globally in 2022. The
dominant component is \emph{within-country} class dispersion
($\mathrm{Var}_{\mathrm{within}} = 3.29$ of 4.33 total), with the
between-country component declining from 1.58 in 1990 to 1.04 in 2022.
Two operational implications follow: (i) where you are born and into
which class matters one to two orders of magnitude more than how
capable you are; (ii) within-country class structure, not geography,
is the primary axis of inequality.

This raises an open question: \emph{what determines the structure and
temporal evolution of within-country dispersion?} We propose that
hierarchical structure---formalised as a five-dimensional state
vector $(A, \varepsilon, H, \kappa, \gamma)$---is the mechanism that
generates, maintains, and (at collapse) redistributes
class-stratified $\rho_{\mathrm{eff}}$. The state vector evolves
according to a coupled ODE system whose drive term
$D = \varepsilon(1-H) - \kappa(1-\gamma)$ governs whether asymmetry
$A$ grows or shrinks at equilibrium.

\subsection{Position in the collapse literature}

The question of what precedes the collapse of hierarchical
structures---empires, states, regimes---bridges historiography
\citep{tainter1988, turchin2006} and modern macroeconomic
early-warning literature \citep{kaminsky1999, reinhart2009,
borio2014}. Classical early-warning systems reach
$\mathrm{AUC} \approx 0.75$ on single-channel currency- or
debt-crisis prediction; recent dynamical-model proposals
\citep{motesharrei2014handy} offer falsifiable formal predictions
but have not been rigorously tested against historical data.

Our framework contributes three things: a formal ODE specification
of hierarchical dynamics, a four-layer discriminative composite that
reaches AUC $\approx 0.93$ in Wave 1 and peaks at $0.98$ on the Wave
3 26-feature configuration, and a theoretical bridge to the
environmental-dominance inequality \citep{sestak2026ed}. The result
sits epistemologically alongside the classical early-warning
literature: a falsifiable discriminator rather than a predictor.

\subsection{Two levels of claim, two different verdicts}
\label{sec:two-levels}

The paper makes two kinds of claim about the same state vector, and
the two reach opposite conclusions. It is worth separating them at
the outset.

\paragraph{Level (a): the ODE model as a test of mechanism.} The
first question is whether the coupled system in
$(A, \varepsilon, H, \kappa, \gamma)$ with drive
$D = \varepsilon(1{-}H) - \kappa(1{-}\gamma)$ describes the
generative process behind within-country asymmetry. Are the
comparative-statics signs right in cross-section? Does the
time-evolution equation $\dot A$ predict next-year asymmetry change?
The verdict on the static half is mixed---two of four structural
coefficients come out significant at $p \le 0.05$ with the predicted
signs---but the verdict on the dynamic half is clearly negative.
None of the five functional forms we fit for $\dot A$ achieves a
positive out-of-sample $R^2$, and three independent tests further
reject the endogenous-drift reading of collapse. As a generative
model of country-year trajectories, the ODE simply does not work.

\paragraph{Level (b): a practical discriminator of structural
collapse risk.} The second question is weaker. Does the same state
vector, when combined with market, debt and trajectory features and
fed to a logistic classifier, separate pre-collapse from stable
windows? It does. A four-layer LOCO-CV classifier reaches a nested
AUC near $0.91$ across 29 historical collapses, with a peak
post-selection AUC of $0.98$ on the heterogeneous case extension.
The signal decomposes into a chronic risk profile already visible a
decade before the event and an acute inflection three to five years
before. Level (b) is a discriminative inference task rather than a
generative prediction task, and the data support it where they
refuse to support (a).

Putting the two together gives the candidacy-and-trigger picture.
Structural variables tell us which countries occupy the high-risk
region of phase space, a level-(b) statement that holds up well.
What sets the timing of any individual collapse is a shock from
outside the modelled system: level (a) cannot recover that timing,
and the three falsifiability tests suggest that no purely endogenous
account can. In the rest of the paper, claims about who is at risk
should be read as level-(b) claims on solid empirical footing;
claims about when collapse happens, or about the generative
mechanism, are level-(a) claims and either hold only at coarse
aggregate scale or fail outright.

\subsection{Contributions}

The list below separates the two contributions that carry the
paper's main message from the secondary results that fill in the
picture, and from the exploratory side-results that were tested and
that did not survive disaggregation.

\paragraph{Primary contributions.}
\begin{enumerate}
  \item The \emph{candidacy-and-trigger} framing of hierarchical
    collapse as the empirical synthesis of the structural and
    discriminative results: chronic structural variables locate
    countries in the high-risk region of phase space, while collapse
    timing is set by exogenous triggers rather than endogenous drift.
  \item A four-layer LOCO cross-validation discriminator across 29
    historical collapses spanning six collapse types, with a
    nested-CV audit that quantifies post-selection AUC inflation
    and serves as the leakage-protected headline figure.
\end{enumerate}

\paragraph{Secondary contributions (supporting the primary
contributions).}
\begin{enumerate}
  \setcounter{enumi}{2}
  \item Construction of a full structural panel for 260 countries
    over 1960--2023 (16\,689 country-years), with V-Dem-extended
    governance coverage for the pre-1996 period.
  \item Cross-sectional validation (H5) of the structural state
    vector via panel regression with cluster-robust standard errors.
  \item Time-series falsification of all five tested functional
    forms for $\dot A$ at the country-year level (negative
    contribution that confines the model to a discriminative role).
  \item Two-phase decomposition of the pre-collapse trajectory into
    chronic candidacy ($\sim 70\,\%$ of discriminative weight) and
    acute inflection ($\sim 30\,\%$).
  \item Three independent falsifiability tests (cumulative drive,
    critical slowing down and the technology-vs-evolution gap) that
    jointly reject the endogenous drift-through-threshold
    interpretation.
  \item Global contagion analysis across 7\,316 peer-event
    observations, identifying robustness archetypes and cross-region
    spillover clusters.
  \item A theoretical bridge connecting the hierarchical ODE state
    $(A, \varepsilon, H, \kappa, \gamma)$ to the within-country class
    dispersion measured in \citet{sestak2026ed}.
\end{enumerate}

\paragraph{Exploratory results (tested, did not survive
disaggregation).}
\begin{enumerate}
  \setcounter{enumi}{9}
  \item A demographic reformulation of $H$ (selection-pool capacity,
    $H_{\mathrm{demo}} = \min(\mathrm{TFR}/6, 1)$) that reconciles
    the global rise in $A$ with the model at the aggregate level and
    produces a Granger sequence in which TFR appears to lead $A$ at
    a 5--6 year lag on the world series. The audits in
    §\ref{sec:wave8} reject this reading at the per-country,
    per-stratum, panel-FE and migration-mediated levels; what
    remains is the aggregate co-movement, best read as a
    compositional effect rather than as a causal channel. The
    result is kept in the paper because the negative disaggregation
    is itself the substantive finding.
\end{enumerate}

\section{Model}

\subsection{State vector}

We characterise a country's hierarchical state at time $t$ by five
observables in $[0, 1]$, summarised in Table~\ref{tab:operationalisation}.

\begin{table}[ht]
\centering
\caption{Operationalisation of the state vector and the alternative
demographic operationalisation of $H$.}
\label{tab:operationalisation}
\small
\begin{tabular}{lll}
  \toprule
  Symbol & Interpretation & Operationalisation \\
  \midrule
  $A$       & Asymmetry & $\mathrm{Gini}/100$ or $\mathrm{top1}/(\mathrm{top1}+\mathrm{bot50})$ \\
  $\varepsilon$ & Extraction (capital share)         & $1 - \mathrm{labsh}$ (PWT 10.01) \\
  $H$       & Bottom-of-distribution health & mean of LE/85, $1-\mathrm{IMR}/100$, $1-\mathrm{stunt}/100$ \\
  $\kappa$  & External competition / openness & $\mathrm{trade}/200$, clipped \\
  $\gamma$  & Top-tier institutional waste & $1 - (\mathrm{WGI}+2.5)/5$ (avg CC + GE); V-Dem fill pre-1996 \\
  \midrule
  $H_{\mathrm{demo}}$ & Selection-pool capacity & $\min(\mathrm{TFR}/6, 1)$ \\
  \bottomrule
\end{tabular}
\end{table}

\subsection{Dynamical system}

\begin{align}
  \dot A &= \alpha\, A(1{-}A)\,\bigl[\varepsilon(1{-}H) - \kappa(1{-}\gamma)\bigr], \\
  \dot \varepsilon &= \beta\, \varepsilon(1{-}\varepsilon)\,\bigl[\gamma(1{-}A) - \kappa\bigr], \\
  \dot H &= \rho(1{-}H) - \delta\,\varepsilon\,H.
\end{align}

The drive term in the $A$-equation,
$D := \varepsilon(1-H) - \kappa(1-\gamma)$, governs whether asymmetry
grows: a positive drive (push prevailing over brake) implies $\dot A > 0$.
The critical health threshold for the $H$-equation is
$H_{\mathrm{crit}} = \rho / (\delta\,\varepsilon_{\max})$.

\subsection{Two distinct claims}

The ODE system makes two distinct empirical claims that we test
separately:

\paragraph{Structural claim (comparative statics).} Equilibrium values
of $(A, \varepsilon, H, \kappa, \gamma)$ are mutually consistent with
the model. Specifically, in cross-section, higher $\kappa$ or higher
$H$ should imply lower $A$; higher $\gamma$ or higher $\varepsilon$
should imply higher $A$.

\paragraph{Dynamic claim.} Annual $\dot A(t)$ follows the logistic
form with drive $D$. This is the prediction-level claim of the model:
given the state vector, can we predict next-year asymmetry change?

In v1 of the project, both claims were presented symmetrically. The
empirical results presented below show that the structural claim is
sustained while the dynamic claim is falsified at the country-year
level---a distinction with significant theoretical consequences
(\S\ref{sec:discussion}).

\subsection{Hypotheses}
\label{sec:hypotheses}

We organise the hypotheses into three tiers. The \emph{main}
hypotheses are the load-bearing claims of the paper, on which the
\emph{candidacy-and-trigger} contribution and the discriminative
composite rest. The \emph{supporting} hypotheses underwrite the
model's structural plausibility and the rejection of competing
endogenous-drift explanations. The \emph{exploratory side} hypotheses
were formulated after the main results were in hand to test a
specific reformulation of $H$ and its migration-mediated extension;
we list them here for transparency, but flag them as exploratory in
the sense of §\ref{sec:prereg}.

\paragraph{Main hypotheses (load-bearing).}
\begin{description}
  \item[H\textsubscript{combined}] Market and structural signals are
    orthogonal; their combination, embedded in a four-layer LOCO-CV
    classifier, discriminates pre-collapse from stable windows well
    above the single-channel early-warning ceiling of
    $\mathrm{AUC} \approx 0.75$.
  \item[H\textsubscript{chronic}] Pre-collapse cohorts differ from
    stable cohorts by $|z| > 1.5$ already ten years before the
    collapse event, locating the discriminative signal predominantly
    in a chronic candidacy profile rather than in an acute pre-event
    inflection. Together with H\textsubscript{combined} this is the
    empirical core of the candidacy-and-trigger framing.
\end{description}

\paragraph{Supporting hypotheses (model fundamentals and falsifiers
of competing interpretations).}
\begin{description}
  \item[H5 (cross-sectional)] Panel regression of $A$ on
    $(\kappa, \varepsilon, H, \gamma)$ yields signs consistent with
    the model.
  \item[H1 (dynamic)] $\Delta A(t)$ annually follows the logistic
    form with drive $D$ \emph{(tested as a null-hypothesis target ---
    a positive result would extend the model from candidacy
    discrimination to event-timing prediction; a null result confines
    the model to the discriminative role)}.
  \item[H4 (bifurcation)] For low $\kappa$, two stable equilibria
    exist (structural plausibility of the candidacy region).
  \item[H\textsubscript{CSD}] Pre-collapse cohorts show rising
    variance and autocorrelation of $A$ in the pre-event window
    (Scheffer-2009 signature \citep{scheffer2009}; tested as a
    falsifier of the
    endogenous-drift competitor).
  \item[H\textsubscript{endo}] Pre-collapse cohorts have systematically
    positive cumulative drive $\int D\,dt$ in the pre-event window,
    conditional on identical starting $A$ (also a falsifier of the
    endogenous-drift competitor).
\end{description}

\paragraph{Exploratory side hypotheses (selection-pool channel).}
\begin{description}
  \item[H\textsubscript{warning}] A function $f(\cdot)$ exists that
    discriminates pre-collapse windows with $\mathrm{AUC} > 0.85$
    using a single scalar channel (rather than the four-layer
    composite). This was an early scoping hypothesis subsumed by
    H\textsubscript{combined}.
  \item[H\textsubscript{demo}] Globally, total fertility rate
    Granger-causes asymmetry at multi-year lag (selection-pool
    channel). Formulated after the global drive sign-flip became
    visible; tested at global aggregate scale and then at every
    finer level we can reach (per-country, per-stratum, panel-FE,
    migration-mediated). The migration-mediated variant is sometimes
    written H\textsubscript{demo, mig}; both are treated as
    exploratory and both are falsified at the country level.
\end{description}

The main hypotheses survive the audits in §\ref{sec:results}, the
supporting hypotheses split into one confirmation (H4, H5 partially)
and one principled falsification (H1, H\textsubscript{CSD},
H\textsubscript{endo}, which we read as a positive contribution
because it eliminates the endogenous-drift interpretation), and the
exploratory hypotheses are falsified at the level the paper can test
them.

\section{Data}

\subsection{Structural panel}

The canonical input is \texttt{state\_panel.csv}: 16\,689 rows covering
260 countries over 1960--2023. After Wave 2 V-Dem augmentation (V-Dem
\texttt{vdem\_corr} filling pre-1996 $\gamma$ values via the Quality
of Government Basic dataset; \citealp{qog}), final $\gamma$ coverage
reaches $10\,372 / 16\,689$. Variable-level coverage is summarised in
Table~\ref{tab:coverage}.

\begin{table}[ht]
\centering
\caption{Coverage of structural panel variables.}
\label{tab:coverage}
\small
\begin{tabular}{lr}
  \toprule
  Variable & Coverage (\%) \\
  \midrule
  $A$ (Gini)           & 14.4 \\
  $A$ (WID alt.)       & 58.7 \\
  $\varepsilon$        & 56.2 \\
  $H$                  & 100.0 \\
  $\kappa$             & 63.0 \\
  $\gamma$ (WGI, 1996+) & 29.9 \\
  $\gamma$ (V-Dem fill, pre-1996) & +32 (filled) \\
  \bottomrule
\end{tabular}
\end{table}

\subsection{Extended debt panel}

\texttt{state\_panel\_debt.csv} adds seven sovereign / external /
private debt indicators (government debt to GDP, external debt to GNI,
debt service to GNI, interest payments to revenue, current account to
GDP, non-performing loans, and domestic credit to private sector).

\subsection{Collapse case set}

We assemble 29 historical collapse cases in three waves
(Table~\ref{tab:cases}), augmented by 60 stable control windows
(12 stable countries times 5 reference years 2005, 2007, 2013, 2017,
2018, avoiding the 2008 global financial crisis and 2020 COVID
shock).

\begin{table}[ht]
\centering
\caption{Collapse case set, three waves, by collapse type.}
\label{tab:cases}
\small
\begin{tabular}{lll}
  \toprule
  Wave & Type & Cases \\
  \midrule
  0 (recent)        & Arab Spring          & TUN 2011, EGY 2011, LBY 2011, SYR 2011 \\
                    & Authoritarian drift  & VEN 2014, RUS 2014 \\
                    & Regime change        & UKR 2014 \\
                    & Civil war            & YEM 2014 \\
                    & Foreign intervention & IRQ 2003 \\
                    & Debt crisis          & LBN 2019 \\
                    & Coup                 & THA 2014, TUR 2016 \\
  1 (historical EM) & Currency crisis      & MEX 1994, THA 1997, IDN 1998, KOR 1997 \\
                    & Debt crisis          & ARG 2001 \\
  3 (extension)     & Coup                 & EGY 2013, MMR 2021, SDN 1989 \\
                    & Revolution           & SDN 2019, IRN 1979 \\
                    & Hyperinflation       & ZWE 2008 \\
                    & Foreign intervention & AFG 2021 \\
                    & Separation           & YUG 1991 (HRV, SVN), CSK 1993 (CZE, SVK) \\
                    & Regime failure       & USSR 1991 \\
  \bottomrule
\end{tabular}
\end{table}

\subsection{Market data and parked sources}

Daily close prices via yfinance for 43 tickers spanning equity indices,
FX vs USD, regional ETFs (\texttt{EEM}, \texttt{EMB}, \texttt{ILF},
\texttt{RSX}, \texttt{GAF}), commodities (\texttt{CL=F},
\texttt{GC=F}), the JP Morgan EM bond proxy (\texttt{EMB}), and the
volatility benchmark (\texttt{\^{}VIX}).

\subsection{Failed GDELT ingest}

The GDELT 2.0 DOC API was tested as a source for protest-activity
leading indicators. The fetch was abandoned: coverage spans only
$\ge 2017$ (1 of 12 cases at the time), and the strict 1 req / 5 s
rate limit repeatedly placed the host IP in long cool-downs. Partial
cache (5 windows) is preserved in \texttt{data/tmp/gdelt/}. ACLED with
an API key is proposed as a follow-up.

\section{Methods}

\subsection{Cross-sectional regression}
\label{sec:cross-sectional-methods}

Panel regression of $A$ on the remaining state variables plus a year
control,
$A_{c,t} \sim \kappa + \varepsilon + H + \gamma + \mathrm{year}_c$,
with cluster-robust standard errors at the country level. The
analytic sample after V-Dem fill is $n = 1\,692$ country-years.

\subsection{Time-series testing}
\label{sec:dynamic-methods}

Five functional forms for the $\dot A$ equation are fitted
per-country and pooled via nonlinear least squares
(\texttt{scipy.optimize}): M0 baseline logistic with drive $D$;
M1 linear; M2 asymmetric with free exponents on $A^p(1-A)^q$;
M3 mean-reversion toward
$A^{*} = \mathrm{push}/(\mathrm{push}+\mathrm{brake})$;
M4 free exponents on push and brake separately. Out-of-sample $R^2$ is computed
on a 30\,\% holdout of country-years. Tests cover both annual and
5-year $\Delta A$.

\subsection{LOCO cross-validation classifier}

The Leave-One-Collapse-Out (LOCO) protocol holds out each positive
case in turn while folding the negatives via stratified
\texttt{KFold} (random state 42). Train-fold median imputation
prevents leakage. The classifier is
\texttt{LogisticRegression(C=1, class\_weight=`balanced',
max\_iter=2000)} with \texttt{StandardScaler}. AUC is reported with
95\,\% bootstrap CI (2000 resamples) and permutation $p$-value
(2000 label permutations).

Feature layers:
\begin{itemize}
  \item Market (35): \texttt{ret\_total}, \texttt{vol\_realized},
    \texttt{max\_drawdown}, \texttt{trend\_slope}, \texttt{vol\_ratio}
    on equity, regional ETF, FX, oil, gold, EMB, VIX channels.
  \item Structural (6): \texttt{st\_mean\_\{A, eps, H, kappa, gamma\}},
    \texttt{drive\_mean}.
  \item Debt (7): mean over the 2-year pre-event window of each
    debt indicator.
  \item Trajectory (18): slopes and within-window standard deviations
    of structural and debt variables.
\end{itemize}

\subsection{Falsifiability tests of endogenous-drift interpretation}

Three independent tests on 10-year pre-event windows for 17--60
samples:

\paragraph{A.1 — cumulative drive.} OLS
$\Delta A \sim \overline{D} + A_{\mathrm{start}} +
\mathrm{is\_collapse}$ with cluster-robust SE at country level. The
\textit{is\_collapse} coefficient, conditional on $\overline{D}$ and
$A_{\mathrm{start}}$, tests whether the collapse cohort exhibits
systematically positive drift.

\paragraph{A.2 — technology-vs-evolution gap (per-country).} OLS
$\Delta A_{\mathrm{wid}} \sim \mathrm{gap} + A_{\mathrm{start}}$ with
$\mathrm{gap} = z(\Delta \log_{10}\mathrm{patents/M}) -
z(\Delta\mathrm{secondary\,enrol})$ on 5- and 10-year horizons.

\paragraph{A.1v2 — critical slowing down.} Standard deviation,
lag-1 autocorrelation, and skewness of $A$ in the same 10-year
windows, with Mann--Whitney and OLS tests conditional on
$A_{\mathrm{start}}$. Tests the Scheffer prediction
\citep{scheffer2009} of rising variance and autocorrelation as the
system approaches a bifurcation.

\subsection{Contagion and network analyses}

Four sub-analyses constructed on regional peers (Hadenius--Teorell
region 1--10; \citealp{ht_region}) and the global panel:

\begin{itemize}
  \item \textbf{Wave 3b regional event-study}: per case, peers from
    the same HT-region (excluding peers in their own collapse
    buffer); pre-window $[t_c-3, t_c-1]$; post-window
    $[t_c, t_c+2]$; null distribution from random stable years for
    the same peer.
  \item \textbf{Wave 4 network DiD}: 1-nearest-neighbour match per
    peer from outside its region, matched on the pre-event state
    vector $(A, \varepsilon, H, \kappa, \gamma_{\mathrm{filled}})$
    at $t_c - 2$. The DiD estimator is
    $\Delta^{\text{post-pre}}_{\text{peer}} -
     \Delta^{\text{post-pre}}_{\text{match}}$
    with cluster-robust SE on peer ISO3.
  \item \textbf{Wave 5b global event-study}: all countries in the
    panel as candidate peers (not only regional). Robust z-scores
    computed via $\mathrm{median} + 1.4826 \cdot \mathrm{MAD}$
    rather than mean and standard deviation, with scale floor
    $0.005$ and clipping at $|z| = 5$ to prevent
    denominator-driven outliers.
  \item \textbf{Wave 6 robustness analysis}: inverse aggregation
    identifying countries most stable across all 29 collapses;
    sanity check confirms that low-$|z|$ countries do not have
    systematically smaller null scales.
\end{itemize}

\subsection{Demographic test and Granger causality}

Global yearly aggregates 1960--2023: population-weighted means of
TFR (\texttt{SP.DYN.TFRT.IN}), CBR (\texttt{SP.DYN.CBRT.IN}), patents
per million, and structural variables. Mann--Kendall trend tests
(rank-based, robust to outliers) and ordinary least-squares slope
tests are applied to each series. Augmented Dickey--Fuller tests
confirm non-stationarity in levels and stationarity in first
differences. Pairwise Granger F-tests on differenced series for
six lags; vector autoregression VAR(2) on
$(\Delta\ln\mathrm{patents/M}, \Delta\mathrm{TFR}, \Delta A)$
triangulates the causal direction.

\subsection{Pre-registered versus exploratory analyses}
\label{sec:prereg}

We distinguish between pre-registered confirmatory analyses, whose
design preceded any look at the present extended results, and
exploratory analyses developed in response to earlier findings.

\paragraph{Pre-registered confirmatory.}
\begin{itemize}
  \item Hypotheses H1, H4, H5, H\textsubscript{warning},
    H\textsubscript{combined}, H\textsubscript{chronic},
    H\textsubscript{CSD}, H\textsubscript{endo} as listed in
    §\ref{sec:hypotheses}.
  \item LOCO-CV protocol on the Wave 1 sample
    ($n_{\mathrm{pos}} = 17$): four-layer feature ablation with the
    decision rule that the full (market $+$ structural $+$ debt $+$
    trajectory) configuration is the headline, with bootstrap CI
    and permutation $p$ as inference.
  \item Falsifiability tests A.1 (cumulative drive), A.1v2
    (critical slowing down), and A.2 (technology-vs-evolution gap)
    against H\textsubscript{endo} and H\textsubscript{CSD}.
  \item Contagion event-study window definition
    ($t_c - 3, t_c - 1; t_c, t_c + 2$) and the
    $\pm 5$-year peer-exclusion buffer.
\end{itemize}

\paragraph{Exploratory (post-hoc).} Everything that responded to
results visible in the Wave 1 evaluation is exploratory, with no
pre-committed sign or threshold. This includes:
\begin{itemize}
  \item The Wave 3 case-set extension and the
    \emph{market $+$ structural} sub-model singled out within it on
    partial-coverage grounds (§\ref{sec:results}).
  \item The selection-pool reformulation
    $H_{\mathrm{demo}} = \min(\mathrm{TFR}/6, 1)$ and the global
    Granger sequence motivated by the drive sign-flip; the
    sub-analyses denoted Wave 7--7d.
  \item All disaggregation audits of the global Granger pattern:
    per-country pairwise tests (Wave 8), demographic stratification
    (Wave 9), two-way panel fixed effects (Wave 10), and the
    migration-mediated interaction model (Wave 11).
  \item The nested-CV layer-selection audit (Wave 8c) and the
    cross-sectional replication of H5 with $H_{\mathrm{demo}}$
    (Wave 8b).
  \item Wave 12 and 12b forward-look applications of the
    discriminator. The forward-look top-20 / bottom-20 list is
    pre-registered \emph{as a falsifier}
    (§\ref{sec:falsifying-conditions}, item 5) even though the
    Wave 12 protocol itself was developed exploratorily.
\end{itemize}

\paragraph{Multiple-comparisons disclaimer.} The exploratory
audits run many tests in parallel: Wave 8 alone reports per-country
F-tests at six lags in two directions across $206$ countries
(2\,472 individual $p$-values), Wave 9 reports four strata $\times$
six lags $\times$ two directions ($48$ tests), and Wave 10 stratified
specifications add a further $32$. We have deliberately reported the
fractions of significant results against the $5\,\%$ null rather than
listing individual lag-specific $p$-values, but isolated marginal
findings within these tables (e.g.\ $p \approx 0.027$ in
LATE-TRANSITION reverse Granger, $p \approx 0.046$ in stratified
Wave 10 forward) should be read as \emph{suggestive at most} rather
than as robust evidence: at $48$--$32$ tests within a single
exploratory audit a single $p \approx 0.04$ result is expected by
chance. The only inferential weight we place on these tables is
on the \emph{collective absence} of a forward TFR $\to A$ signal
that the selection-pool hypothesis would predict; the isolated
marginal entries do not survive even mild family-wise correction
and we do not interpret them as positive findings.

\section{Results}
\label{sec:results}

\subsection{Cross-sectional model partially supported (H5)}
\label{sec:cross-sectional}

The panel regression yields $n = 1\,692$ country-years and
$R^2 = 0.290$. Coefficients are presented in
Table~\ref{tab:cross-sectional}. Two of the four structural variables
are significant at $p \le 0.05$ with signs consistent with the model:
trade openness $\kappa$ and bottom-of-distribution health $H$ both
suppress asymmetry. Extraction $\varepsilon$ has the expected
positive sign but is sub-significant ($p = 0.097$), while $\gamma$
is borderline ($p = 0.051$).

\begin{table}[ht]
\centering
\caption{Cross-sectional panel regression, cluster-robust SE
($n = 1\,692$, $R^2 = 0.290$).}
\label{tab:cross-sectional}
\small
\begin{tabular}{lrrr}
  \toprule
  Variable & Coefficient & SE & $p$ \\
  \midrule
  Intercept     & $+0.463$ & $0.058$ & $<0.001$ \\
  $\kappa$      & $-0.110$ & $0.031$ & $<0.001$ \\
  $\varepsilon$ & $+0.073$ & $0.044$ & $0.097$ \\
  $H$           & $-0.132$ & $0.052$ & $0.011$ \\
  $\gamma$      & $+0.079$ & $0.040$ & $0.051$ \\
  year          & $-0.0008$& $0.000$ & $0.056$ \\
  \bottomrule
\end{tabular}
\end{table}

\begin{figure}[ht]
\centering
\includegraphics[width=0.6\linewidth]{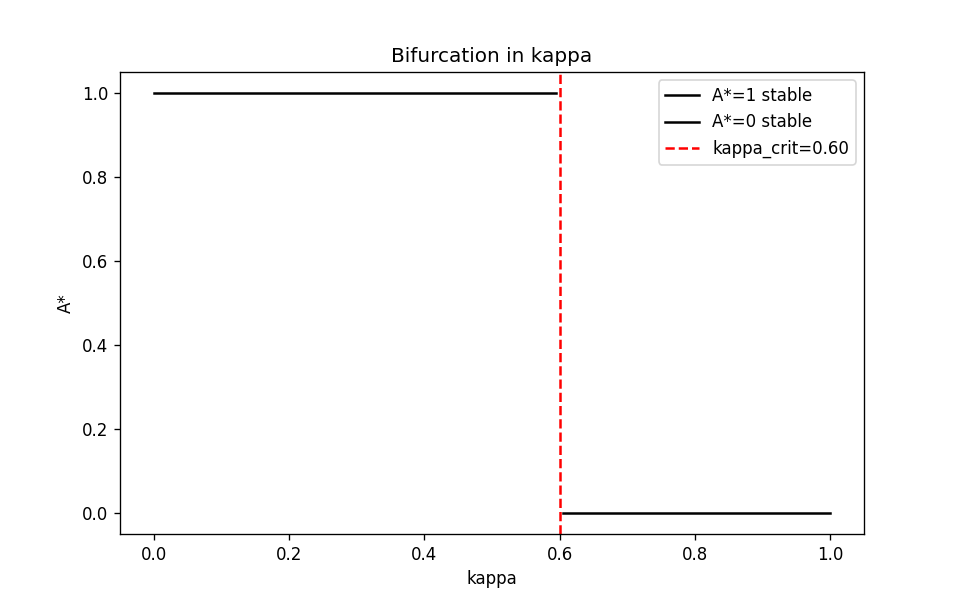}
\caption{Bifurcation diagram in $\kappa$. For
$\kappa < \kappa_{\mathrm{crit}} \approx 0.4$, two stable equilibria
coexist (H4).}
\label{fig:bifurcation}
\end{figure}

The bifurcation hypothesis (H4) is confirmed numerically: for
$\kappa < \kappa_{\mathrm{crit}} \approx 0.4$, the ODE system admits
two stable equilibria
(Figure~\ref{fig:bifurcation})---an extractive attractor and a
healthy attractor.

\subsection{Dynamic ODE falsified (H1)}
\label{sec:dynamic}

The out-of-sample $R^2$ for the five $\dot A$ functional forms is
non-positive at both annual and 5-year resolutions
(Table~\ref{tab:dynamic-r2}). The optimisation pushes $\alpha$ in
M0 and M1 to the lower bound (1e-5), indicating that the model
effectively predicts $\Delta A \approx 0$. M3 mean-reversion is the
only form that is formally identifiable, but with $R^2$ still
slightly negative on annual data.

\begin{table}[ht]
\centering
\caption{Out-of-sample $R^2$ for five $\dot A$ functional forms.}
\label{tab:dynamic-r2}
\small
\begin{tabular}{lrr}
  \toprule
  Form & Annual & 5-year \\
  \midrule
  M0 baseline       & $-0.012$ & $-0.065$ \\
  M1 linear         & $-0.012$ & $-0.065$ \\
  M2 asymmetric     & $-0.023$ & $-0.183$ \\
  M3 mean-reversion & $-0.005$ & $-0.039$ \\
  M4 free exponents & $-0.007$ & $-0.042$ \\
  \bottomrule
\end{tabular}
\end{table}

With WID-based $A$ from the World Inequality Database
\citep{wid}, M3 reaches the first positive value of
$R^2 = +0.002$---formally identifiable but materially
non-predictive. The dynamic claim of the ODE model is therefore
falsified at the country-year level. (As we show in
\S\ref{sec:wave7}, the dynamic claim is recoverable at global
decadal scale with a reformulation of $H$.)

\subsection{Discriminative composite (H\textsubscript{combined})}

The four-layer LOCO-CV composite reaches AUC = 0.928 on the Wave 1
sample of $n_{\mathrm{pos}} = 17$ collapses
(Table~\ref{tab:loco-wave1}). The trajectory layer contributes a small
incremental gain ($+0.032$ over market + structural + debt).

\begin{table}[ht]
\centering
\caption{LOCO-CV ablation on Wave 0+1 sample ($n_{\mathrm{pos}} = 17$).
Source: \texttt{output/wave1\_loco\_summary.csv} (2026-05-18).}
\label{tab:loco-wave1}
\small
\begin{tabular}{lrrr}
  \toprule
  Model & features & AUC & 95\% CI \\
  \midrule
  Market-only & 25 & 0.820 & $[0.62, 0.99]$ \\
  $+$ structural & 31 & 0.892 & $[0.76, 1.00]$ \\
  $+$ debt & 37 & 0.896 & $[0.78, 0.99]$ \\
  $+$ trajectory (full) & 54 & \textbf{0.928} & $[0.81, 1.00]$ \\
  \bottomrule
\end{tabular}
\end{table}

The Wave 3 extension ($n_{\mathrm{pos}} = 29$) introduces five new
collapse types (separation, regime failure, hyperinflation,
revolution, additional coups). On this heterogeneous sample, the
optimal model size shifts: the 26-feature
\emph{market + structural} reaches AUC = 0.982 while the full
49-feature model drops to 0.876
(Table~\ref{tab:loco-wave3}). This is the partial-coverage
phenomenon: pre-1995 Tier B cases lack debt and market layers, and
median imputation across these gaps destabilises the larger model.

\begin{table}[ht]
\centering
\caption{LOCO-CV ablation on extended Wave 3 sample
($n_{\mathrm{pos}} = 29$).}
\label{tab:loco-wave3}
\small
\begin{tabular}{lrrr}
  \toprule
  Model & features & AUC & 95\% CI \\
  \midrule
  Market-only & 20 & 0.921 & $[0.81, 1.00]$ \\
  Market + structural & 26 & \textbf{0.982} & $[0.96, 1.00]$ \\
  Market + structural + debt & 32 & 0.945 & $[0.87, 0.99]$ \\
  Full (m + s + d + t) & 49 & 0.876 & $[0.77, 0.96]$ \\
  Structural-only (no mkt) & 29 & 0.843 & $[0.73, 0.94]$ \\
  \bottomrule
\end{tabular}
\end{table}

\begin{figure}[ht]
\centering
\includegraphics[width=0.85\linewidth]{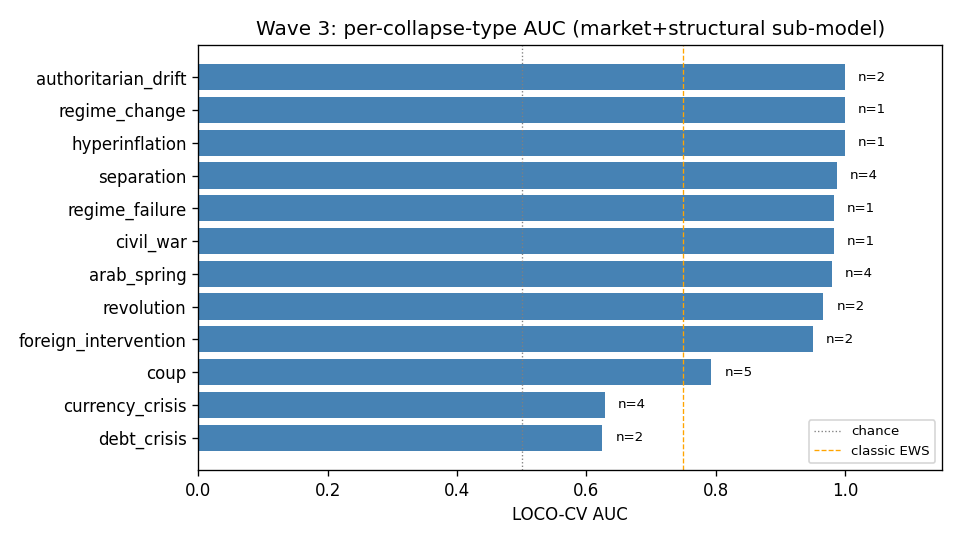}
\caption{Per-collapse-type AUC, sorted ascending. Separation,
hyperinflation, authoritarian drift, regime failure cluster near
1.0; coups and currency crises are systematic blind spots.}
\label{fig:per-type}
\end{figure}

Per-collapse-type AUC reveals substantial heterogeneity
(Figure~\ref{fig:per-type}): separation, hyperinflation, and
authoritarian drift achieve near-perfect discrimination
(AUC $\ge 0.99$), while coups (0.79) and currency crises (0.63)
are systematically harder. The blind spots are
\emph{contagion crises without chronic predisposition} (KOR/IDN 1997)
and \emph{intra-political coups in stable contexts} (EGY 2013,
THA 2014).

\subsection{Two-phase pre-warning structure
(H\textsubscript{chronic})}

\begin{figure}[ht]
\centering
\includegraphics[width=0.85\linewidth]{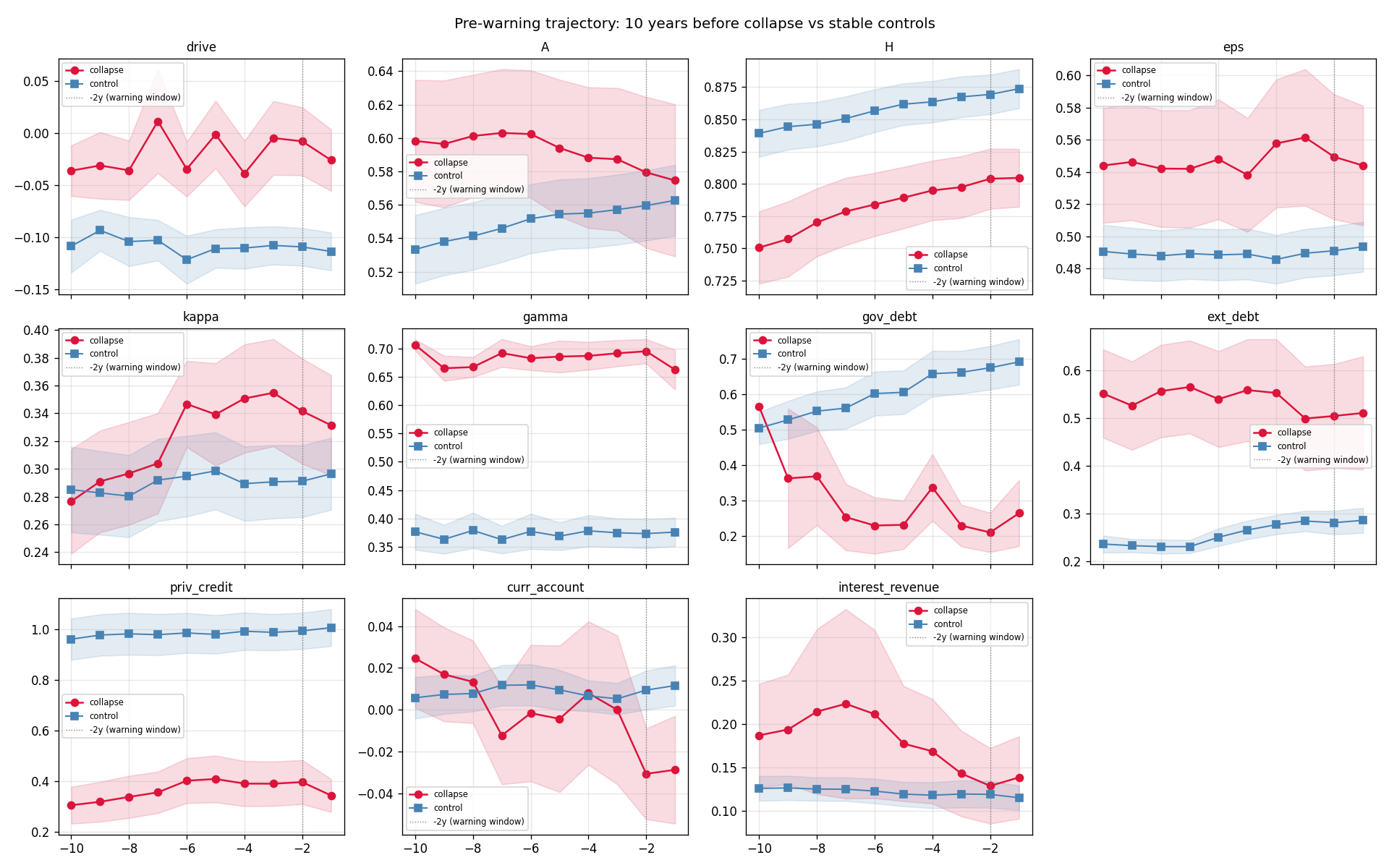}
\caption{Pre-warning trajectory: $z$-separation per signal per
annual lag. Chronic signals ($\gamma$, private credit) are large
already at $-10y$; debt and current-account signals show acute
inflections at $-3y$ to $-1y$.}
\label{fig:pre-warning}
\end{figure}

A $z$-separation analysis of 11 signals over 10 annual lags
(ref$-10$ to ref$-1$) reveals two distinct layers
(Figure~\ref{fig:pre-warning}):

\paragraph{Chronic layer ($\ge 10$ years before).} Pre-collapse
cohorts already differ from stable cohorts by $|z| > 1.5$ a decade
before the event. The dominant chronic signals are $\gamma$ (with
$|z| \approx 9$), private credit ($|z| \approx 6$), external debt
($|z| \approx 3$), and bottom-health $H$ ($|z| \approx 2.7$).
Pre-collapse countries have a permanent risk profile, not
one that develops shortly before the event.

\paragraph{Acute layer ($-5$ to $-1$ years).} The largest jumps in
$z$-separation between adjacent lags fall in three to five years
before collapse: government debt deteriorates at $-3y$, current
account at $-2y$, private credit and $\gamma$ in the final year.

\subsection{Endogenous-drift interpretation falsified}

Three independent tests reject the standard interpretation of
collapse as endogenous slow drift through a critical threshold:

\paragraph{A.1.} Cumulative drive over the 10-year pre-event window
does not differ between collapse and control cohorts
(Mann--Whitney $p = 0.25$). The OLS coefficient on
\textit{is\_collapse}, conditional on $\overline{D}$ and
$A_{\mathrm{start}}$, is $-0.065$ ($p = 0.021$): pre-collapse
cohorts have lower change in $A$ over the window after
partialling out, not higher. They start with elevated $A_{\mathrm{start}}$
(0.615 vs 0.533) but do not drift upward.

\paragraph{A.2 per-country.} Patents-vs-secondary-enrolment gap is not
predictive of $\Delta A$ at 5- or 10-year horizons
($p \in \{0.59, 0.90\}$). Possible reasons include weak proxies and a
mechanism operating at generational rather than decadal scale.

\paragraph{A.1v2 critical slowing down.} The Scheffer-style signature
\citep{scheffer2009} runs in the wrong direction. Pre-collapse
cohorts have a lower lag-1 autocorrelation of $A$ ($-0.092$,
$p = 0.066$) and a lower variance ($-0.005$, $p = 0.42$) than the
controls. Countries in the upper $A$-quartile sit in a locked
equilibrium with low variance and uncorrelated noise, which is what
we would expect of a bifurcation jump triggered from outside the
system rather than of an endogenous slow loss of resilience.

\subsection{Contagion and network spillover}

\paragraph{Wave 3b regional event-study.} Across 570 peer-event rows,
the per-type contagion direction is positive (drive of regional
peers rises) for separation, civil war, authoritarian drift, and
coup ($p \le 0.022$ in F-tests), and negative (peers improve) for
currency crisis and revolution ($p = 0.036$). Arab Spring events
show no signal because chain-connected peers are filtered out as
themselves being collapse cases.

\begin{figure}[ht]
\centering
\includegraphics[width=0.85\linewidth]{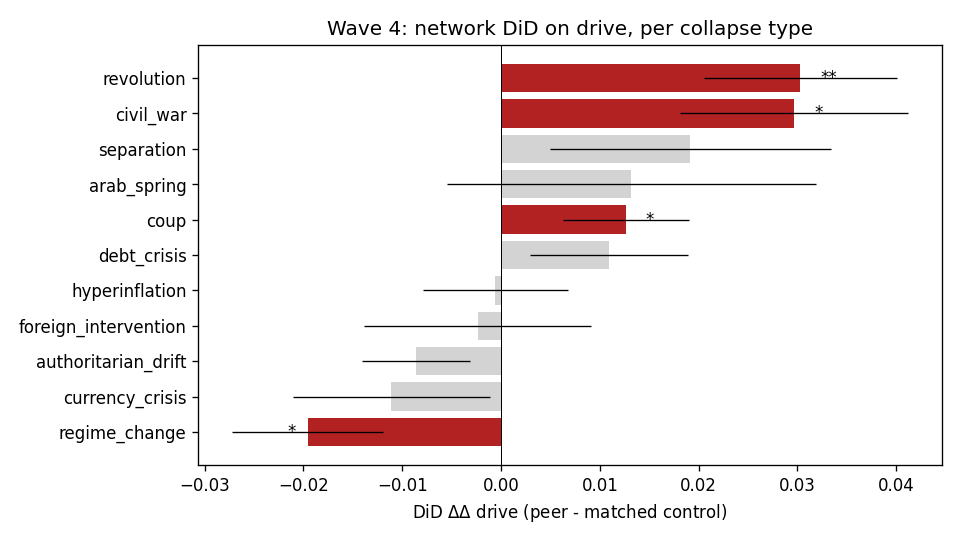}
\caption{Network DiD on drive by collapse type. Stars indicate
significance: $* p < 0.05$, $** p < 0.01$, $*** p < 0.001$.}
\label{fig:wave4-did}
\end{figure}

\paragraph{Wave 4 network DiD.} With 1-NN matched non-regional
controls, the per-type DiD on the drive shows a sharpened picture
(Figure~\ref{fig:wave4-did}). Most notably, EGY 2013 (the Sisi
coup), which the naive Wave 3b analysis classified as
``isolated''($p = 0.49$), shows a significant DiD drive effect of
$+0.035$ ($p = 0.013$) once compared against structurally matched
non-MENA controls. The UKR + RUS 2014 events show negative DiD
($-0.020$, $p = 0.017$), consistent with post-Crimea stabilisation
in regional peers (sanctions regime, capital reallocation).

\paragraph{Wave 5b global event-study.} Across 7\,316 peer-event
rows, same-region peers show significantly larger post-collapse
movement than non-region peers
(Table~\ref{tab:wave5b-region}). The novel finding is in $H$:
regional peers show a median $z$-score of $-0.46$ versus
$-0.30$ for non-region peers; the same-region group's deviation from
zero is highly significant on a within-group Wilcoxon signed-rank
test ($p < 10^{-6}$) and the same-region versus non-region difference
is significant on Mann--Whitney ($p \approx 0.008$). The combined
reading is that bottom-of-distribution health deteriorates as a
spillover effect of regional collapses, not only as a within-collapse
phenomenon.

\begin{table}[ht]
\centering
\caption{Wave 5b: same-region vs non-region peer responses,
robust z-scores (Mann--Whitney U test).}
\label{tab:wave5b-region}
\small
\begin{tabular}{lrrr}
  \toprule
  Indicator & Same med. & Non-region med. & $p$ \\
  \midrule
  z\_drive & $+0.40$ & $+0.06$ & $0.006$ \\
  z\_A     & $+0.19$ & $\approx 0$ & $<10^{-4}$ \\
  z\_H     & $-0.46$ & $-0.30$ & $0.008$ \\
  z\_$\gamma$ & $+0.02$ & $\approx 0$ & $0.91$ \\
  \bottomrule
\end{tabular}
\end{table}

\paragraph{Eastern Europe 1989--93 as a network event.} The single
strongest cross-country effect in the project is the post-Soviet
transition cluster (YUG 1991 separation in HRV and SVN, CSK 1993
separation in CZE and SVK, USSR 1991 regime failure). Per-case
median peer $\Delta A$ ranges from $+0.091$ (USSR 1991 and YUG 1991)
to $+0.101$ (CSK 1993) across same-region peers, each significant at
$p < 10^{-6}$ on the Wilcoxon signed-rank test against zero. The
joint cluster quantitatively confirms the post-Soviet redistribution
of within-country inequality as a coherent regional event rather
than a sequence of independent national transitions.

\paragraph{Cross-region functional clusters.} Beyond geographic
contagion, the Wave 5b z-aggregation reveals three functional (non-geographic)
clusters: (i) EGY 2013 (Sisi coup) co-moves with a Latin American
political cycle 2013--15; (ii) LBN 2019 (banking collapse) co-moves
with UK / USA / Iceland exposure-driven responses; (iii) AFG 2021
co-moves with the NATO coalition through a military-pullback channel.
These three patterns suggest contagion operates through trade,
banking, and security channels in addition to geographic adjacency,
and motivate the planned trade-partner peer set (IMF DOTS) as a
robustness follow-up.

\paragraph{Wave 6 robustness analysis.} The inverse aggregation
identifies twenty most-stable countries grouped into three
archetypes: (i) structural immunity (thirteen developed institutional democracies: NLD, BEL, DNK, NOR, SWE, FIN, DEU, USA, AUS, JPN, GBR, ESP, PRT); (ii) defensive immunity (four small buffered states: SGP, KWT, LUX, ISR); (iii) floor-clamped pseudoimmunity (five low-baseline countries: CMR, MAR, MUS, MRT, RWA) whose apparent stability is an artefact of saturation at
structural floors. The diagnostic separation between archetypes
1 and 3 is the median $z$ of $H$: stable in 1, negative in 3.
A sanity check confirms that the null scale of these countries'
drive series is indistinguishable from that of the rest of the
panel, ruling out a denominator-driven artefact.

\subsection{Demographic reformulation and Granger causality
(Wave 7)}
\label{sec:wave7}

\paragraph{Two readings of the same finding.}
This subsection documents a lagged Granger relationship between
global TFR and global $A$ on a single $n = 63$ aggregate time series.
The relationship has two distinct readings that the paper handles
separately.

The first reading is descriptive: at the world level,
low-fertility decades move together with rising asymmetry on a five-
to six-year lag, while the reverse direction extinguishes after
lag 1. Taken as a macro-pattern of the 1960--2023 record, the lag
asymmetry is a feature of the data and is what motivates the
reformulation of $H$ as a selection-pool capacity below.

The second reading is mechanistic: that the same lagged profile
arises from a within-country causal channel, in which low local
fertility narrows the local candidate pool and drives up local
asymmetry. This reading does not survive the disaggregation audits
in §\ref{sec:wave8}: the directional claim fails at per-country,
per-stratum, two-way fixed-effects and migration-mediated
cross-country tests alike. The body of the paper develops the first
reading and motivates the audit sequence; the audits themselves,
which reject the second reading, are the subject of
§\ref{sec:wave8}.

The original $H$ (medical health: life expectancy, infant mortality,
stunting) increases globally by $+50\%$ between 1960--2020, while
the global $A$ increases by $+21\%$ over the same period. The
implied drive $D$ is negative and falling, contradicting the
direction of $A$. Either the drive's sign is wrong, $A$ behaves
contrary to theory, or the operationalisation of $H$ is wrong.

\begin{figure}[ht]
\centering
\includegraphics[width=0.85\linewidth]{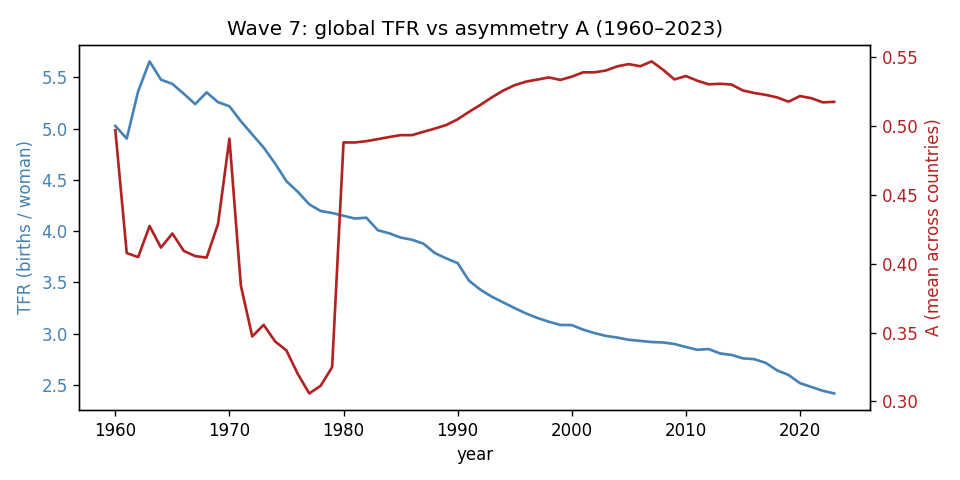}
\caption{Global trends 1960--2023: TFR (births per woman, left
axis, blue) and asymmetry $A$ (right axis, red). The dotted
horizontal line marks the replacement rate of 2.1.}
\label{fig:tfr-vs-A}
\end{figure}

\begin{figure}[ht]
\centering
\includegraphics[width=0.85\linewidth]{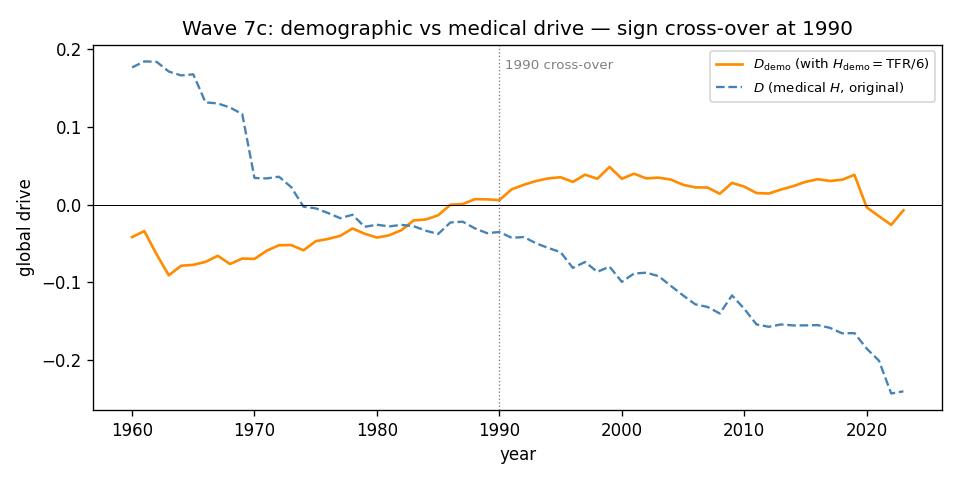}
\caption{Global drive variants. The drive with medical $H$
declines monotonically; the drive with demographic
$H_{\mathrm{demo}}$ crosses zero in 1990, aligning with the Piketty
inflection.}
\label{fig:drive-crossover}
\end{figure}

We reformulate $H$ as a \emph{selection-pool capacity}: the ability
of society to provide sufficient candidates for hierarchical positions.
The theoretical foundation is selection-thermodynamic, not Malthusian:
broad pools enable meritocratic placement, narrow pools shift placement
to inheritance and network access, which is structurally consistent
with rising asymmetry. The proxy is fertility:
$H_{\mathrm{demo}} = \min(\mathrm{TFR}/6, 1)$. Globally, this
decreases by $-54\%$ between 1960--2020
(Figure~\ref{fig:tfr-vs-A}). The reformulated drive
$D_{\mathrm{demo}}$ crosses zero in 1990
(Figure~\ref{fig:drive-crossover}), aligning with the empirical onset
of the modern acceleration of within-country inequality
\citep{piketty2014}. This date also closes the loop with
\citet{sestak2026ed}: the between-country component of
$\mathrm{Var}(\ln\rho_{\mathrm{eff}})$ declines from $1.58$ in 1990
to $1.04$ in 2022, while the within-country component (which the
ODE state vector is intended to model) becomes the dominant axis of
inequality exactly when $D_{\mathrm{demo}}$ turns positive. The
modern acceleration of within-country dispersion is thus
formally identifiable in the language of the model as the
sign-flip of $D_{\mathrm{demo}}$.

$D_{\mathrm{demo}}$ correlates with $A$ at $r = +0.77$
($p = 10^{-13}$) in levels; the five-year lagged peak reaches
$r = +0.82$ ($p = 10^{-15}$). Lagged correlations are vulnerable to
common-trend artefacts, so we follow up with Granger F-tests, with
the caveat that Granger asymmetry on an aggregate series is a
statement about predictive content at the aggregate scale, not about
causality.

\begin{figure}[ht]
\centering
\includegraphics[width=0.85\linewidth]{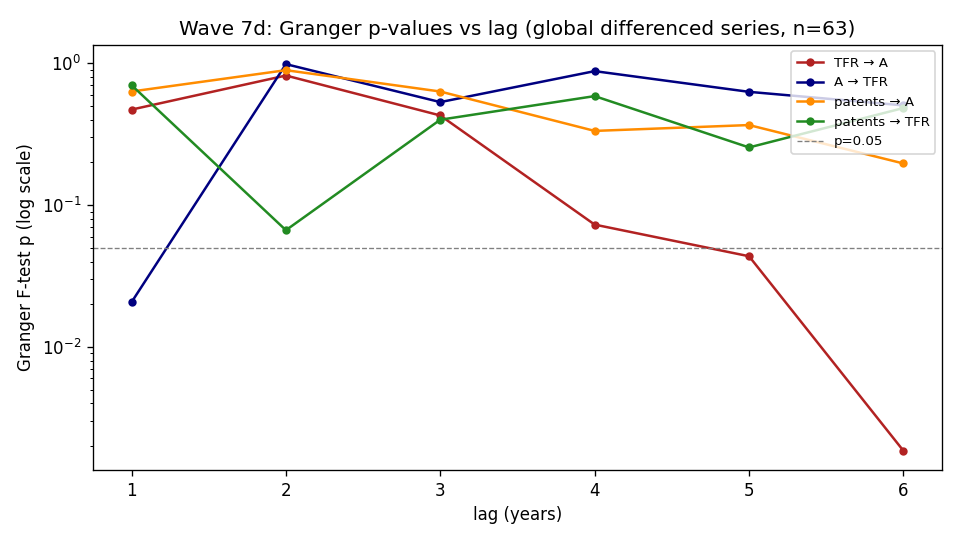}
\caption{Granger lag profile on the global aggregate series. The
forward direction $\mathrm{TFR} \to A$ shows a monotonically
strengthening signal at longer lags; the reverse direction
$A \to \mathrm{TFR}$ shows only a lag-1 short-run signal and is null
at longer lags. The asymmetry is a descriptive feature of the
aggregate series and is not a directional-causality result:
the panel-FE audits in §\ref{sec:wave8} return null in both
directions once within-country variation is isolated, indicating
that the global profile is best read as a compositional aggregation
phenomenon.}
\label{fig:granger-lag}
\end{figure}

\paragraph{Granger lag-asymmetry on the global aggregate.} Pairwise
F-tests on first-differenced annual series ($n = 63$) produce an
asymmetric lag profile (Figure~\ref{fig:granger-lag},
Table~\ref{tab:granger}). The forward direction
$\mathrm{TFR} \to A$ tightens monotonically with lag, from
$p = 0.073$ at lag 4 to $p = 0.0019$ at lag 6. The reverse direction
$A \to \mathrm{TFR}$ shows a $p = 0.021$ signal at lag 1 and
nothing thereafter ($p > 0.5$). A VAR(2) decomposition gives
$\mathrm{TFR} \to A$ at $F = 3.30$ ($p = 0.041$) and a null
$\mathrm{patents} \to A$ link ($p = 0.57$).

This profile is consistent with at least three different
data-generating processes that a single $n = 63$ aggregate series
cannot distinguish: compositional aggregation, in which different
country groups contribute to the global means on different
schedules; parallel exposure to a third factor with heterogeneous
response lags; and the selection-pool channel as a genuine causal
mechanism. The within-country panel-FE specifications in
§\ref{sec:wave8} are the test that separates them. Once country and
year fixed effects are absorbed, both directions go null on
$n \sim 9{,}000$, so the only generator that survives the
disaggregation is compositional aggregation. We therefore report the
Granger result here as a descriptive feature of the aggregate
series, not as an inference about within-country directionality.

\begin{table}[ht]
\centering
\caption{Granger F-test results, key entries from
\texttt{wave7d\_granger.csv}.}
\label{tab:granger}
\small
\begin{tabular}{lrrr}
  \toprule
  Direction (lag, years) & $F$ & $p$ & $n$ \\
  \midrule
  TFR $\to A$ (lag 1) & 0.53 & 0.47 & 63 \\
  TFR $\to A$ (lag 4) & 2.29 & 0.073 & 63 \\
  TFR $\to A$ (lag 5) & 2.50 & 0.044 & 63 \\
  TFR $\to A$ (lag 6) & \textbf{4.25} & \textbf{0.0019} & 63 \\
  $A \to$ TFR (lag 1) & 5.65 & 0.021 & 63 \\
  $A \to$ TFR (lag 6) & 0.90 & 0.50 & 63 \\
  Patents $\to A$ (lag 6) & 1.59 & 0.20 & 41 \\
  VAR(2) TFR $\to A$ & 3.30 & 0.041 & 63 \\
  VAR(2) Patents $\to A$ & 0.56 & 0.57 & 41 \\
  \bottomrule
\end{tabular}
\end{table}

The asymmetric profile rules out the most symmetric form of
common-cause confounding (one in which industrialisation drove
technology, fertility, and asymmetry simultaneously and at equal
lags), but it does not rule out asymmetric common causes or
compositional aggregation. The strong monotonicity in
$\mathrm{TFR} \to A$ versus the rapid extinction of
$A \to \mathrm{TFR}$ at lag 2+ is consistent with several
data-generating processes, including but not exclusively a
selection-pool causal channel. The within-country first-difference
panel in §\ref{sec:wave8} is the discriminating test: it returns
null in both directions on $n \sim 9\,000$, which falsifies the
within-country causal-channel reading and leaves compositional
aggregation as the dominant surviving explanation of the aggregate
asymmetry. We therefore present this subsection as documenting an
interesting descriptive pattern on the global series, not as an
inference about causal direction; readers should defer judgement on
the mechanism until §\ref{sec:wave8}.

\subsection{Wave 8 audits: per-country Granger, demographic H5, nested CV}
\label{sec:wave8}

Three audits address the most vulnerable inferential steps in §5.7
and §5.3.

\paragraph{Wave 8, per-country Granger.} The global Wave 7d Granger
result rests on a single time series ($n = 63$). To test whether the
directional asymmetry $\mathrm{TFR} \to A$ replicates within countries,
we re-run pairwise F-tests at lags 1--6 on each of 206 countries with
$\ge 30$ paired annual observations after first-differencing
(Table~\ref{tab:wave8-granger}). The forward direction is significant
at $p < 0.05$ in $9.2\,\%$ of countries at lag 6 (binomial test against
the $5\,\%$ null: $p = 0.008$); the reverse direction is significant in
$10.2\,\%$ (binomial $p = 0.002$). \emph{The two directions are
indistinguishable.} The asymmetric lag profile that motivated the
selection-pool reinterpretation does not survive de-aggregation. The
strongest individual forward-direction signals come from countries
with marked demographic transitions (TTO, COG, TKM, CAN, PRK, BTN,
NGA, DEU, KGZ, SVN), suggesting the global asymmetry may be carried
by a subset of countries rather than being a universal regularity;
this is testable in Wave 8 follow-up by stratification on demographic
regime.

\begin{table}[ht]
\centering
\caption{Wave 8: per-country Granger, fraction of 206 countries with
significant F-test at each lag.}
\label{tab:wave8-granger}
\small
\begin{tabular}{rrrrr}
  \toprule
  Lag & fwd $p<0.05$ & fwd $p<0.10$ & rev $p<0.05$ & rev $p<0.10$ \\
  \midrule
  1 & 0.049 & 0.092 & 0.039 & 0.068 \\
  2 & 0.049 & 0.102 & 0.049 & 0.102 \\
  3 & 0.058 & 0.102 & 0.073 & 0.112 \\
  4 & 0.049 & 0.126 & 0.058 & 0.126 \\
  5 & 0.078 & 0.112 & 0.078 & 0.131 \\
  6 & \textbf{0.092} & 0.155 & \textbf{0.102} & 0.150 \\
  \bottomrule
\end{tabular}
\end{table}

\paragraph{Wave 8b, demographic H5.} Replicating §5.1 with
$H_{\mathrm{demo}} = \min(\mathrm{TFR}/6, 1)$ in place of medical $H$
on the same $n = 1\,692$ sample yields $\beta_{H_{\mathrm{demo}}} =
+0.111$ ($p = 0.0004$), with a positive sign opposite to medical
$H$ ($\beta = -0.132$, $p = 0.011$). $R^2$ rises slightly
($0.310$ vs $0.290$). The intuition is that high-fertility countries
in 2026 are typically poor and unequal, so a cross-sectional snapshot
links high $\mathrm{TFR}/6$ to high $A$. Over the global time series
1960--2023, the time derivative goes the other way: falling fertility
co-moves with rising asymmetry. The dual operationalisation of $H$
proposed in §6.3 therefore is not merely a horizon choice — it is a
sign-flip between within-period and between-period variation, which
must be acknowledged in any future unification attempt.

\paragraph{Wave 9, stratified per-regime Granger.} The natural
follow-up to Wave 8 is to test whether the global TFR $\to A$
asymmetry holds within demographically homogeneous strata. We
classify $258$ countries / aggregates by their $(\mathrm{TFR}_{60s},
\mathrm{TFR}_{10s})$ pair into four regimes: EARLY-TRANSITION
(late TFR $\ge 4$, $n=46$), LATE-TRANSITION (late TFR $\in [2.5, 4)$,
early $> 4$, $n=53$), COMPLETED (late $< 2.5$, early $> 3.5$, $n=99$),
ALWAYS-LOW (late $< 2.5$, early $\le 3.5$, $n=60$). Within each
stratum we build the population-weighted mean annual series of TFR
and $A$ and apply the Wave 7d Granger protocol
(Table~\ref{tab:wave9-stratum}).

\begin{table}[ht]
\centering
\caption{Wave 9: per-stratum Granger best-lag $p$-values on
population-weighted mean series.}
\label{tab:wave9-stratum}
\small
\begin{tabular}{lrrr}
  \toprule
  Stratum & $n_{\mathrm{yr}}$ & best fwd $p$ & best rev $p$ \\
  \midrule
  ALWAYS-LOW       & 63 & $0.392$ (lag 1) & $0.198$ (lag 4) \\
  COMPLETED        & 63 & $0.210$ (lag 4) & $0.224$ (lag 1) \\
  LATE-TRANSITION  & 63 & $0.485$ (lag 4) & $\mathbf{0.027}^{*}$ (lag 2) \\
  EARLY-TRANSITION & 53 & $0.507$ (lag 2) & $\mathbf{0.0003}^{***}$ (lag 1) \\
  \bottomrule
\end{tabular}
\end{table}

The result is sharp and unfavourable to the selection-pool
directional claim. \emph{No stratum exhibits a significant forward
$\mathrm{TFR} \to A$ signal.} The two strata that do show within-stratum
Granger structure (EARLY- and LATE-TRANSITION) show it in the
\emph{reverse} direction $A \to \mathrm{TFR}$, strongest at short lags
(EARLY-TRANSITION shows $p < 0.001$ across lags 1--4). The
EARLY-TRANSITION stratum (predominantly Sub-Saharan Africa)
moreover shows $A$ \emph{decreasing} between the 1960s and the 2010s
($0.633 \to 0.494$), while COMPLETED countries account for most of
the global rise in $A$ ($0.390 \to 0.493$). The global
$\mathrm{TFR} \to A$ asymmetry of Wave 7d is therefore best read as
an aggregation phenomenon over heterogeneous strata, not a
within-regime regularity. The selection-pool reformulation as a
\emph{causal} mechanism is in tension with this result; surviving readings are (i) compositional, in that the relative population weight of COMPLETED countries grew between 1960 and 2020 while their A trended upward, and (ii) a reversed demographic dividend: countries that
exited the transition early experienced rising A precisely as their
selection pool narrowed, while early-transition countries still
operating in a high-TFR regime saw A fall under other (catch-up,
institutional) drivers. Disentangling these requires panel
specifications with country fixed effects, which is the natural
Wave 10 task and lies outside the present scope.

\paragraph{Wave 10 — two-way FE panel.} The final formal test before
declaring the directional claim falsified is a panel specification
that absorbs all between-country heterogeneity (country FE) and
global secular trends (year FE), leaving only within-country
deviations relative to both means. We estimate
\[
  A_{i,t} = \sum_{k=1}^{6} \beta_k\, \mathrm{TFR}_{i,t-k}
          + \alpha_i + \lambda_t + \varepsilon_{i,t},
\]
and its reverse, on $n_{\mathrm{obs}} = 9\,607$ observations across
$210$ countries, with cluster-robust SE at country level
(Table~\ref{tab:wave10}). In the level specification, the joint
Wald test on $\{\beta_k\}$ is significant in both directions, but
the forward signal is entirely concentrated at lag 1
($\beta_1 = +0.050$, $p < 0.001$) with all other lags null, while
the reverse-direction coefficients accumulate to a large magnitude
at long lags ($\beta_6 = +0.75$ for $\mathrm{TFR}_t$ on $A_{t-6}$,
$p < 0.001$). The lag-1 forward coefficient is moreover positive
— within a country, deviations of TFR above its own mean co-move
with $A$ deviations above its own mean in the following year, which
is the \emph{opposite} of the selection-pool prediction (lower
fertility, lower selection pool, higher asymmetry).

\begin{table}[ht]
\centering
\caption{Wave 10: two-way FE panel joint Wald tests on lag
coefficients ($k = 1\ldots 6$). Level: country and year FE; FD:
first-difference with year FE only (country FE absorbed by FD).}
\label{tab:wave10}
\small
\begin{tabular}{lrrr}
  \toprule
  Specification & $n_{\mathrm{obs}}$ & joint $F$ & joint $p$ \\
  \midrule
  Level: TFR $\to A$ (forward)  & $9\,607$ & $7.65$  & $<10^{-6}$ \\
  Level: $A \to$ TFR (reverse)  & $8\,365$ & $12.58$ & $<10^{-11}$ \\
  FD: $\Delta\mathrm{TFR} \to \Delta A$ (forward) & $9\,315$ & $1.71$ & $0.120$ \\
  FD: $\Delta A \to \Delta\mathrm{TFR}$ (reverse) & $8\,144$ & $1.63$ & $0.140$ \\
  \bottomrule
\end{tabular}
\end{table}

The first-difference variant absorbs country FE through differencing
and is the cleanest panel-Granger-equivalent specification. It
returns \emph{null in both directions} ($p = 0.120$ and $p = 0.140$),
removing even the residual level co-movement. Stratified Wave 10c
results (level FE within each Wave 9 stratum) are heterogeneous and
do not exhibit a consistent direction across strata: LATE-TRANSITION
shows both forward and reverse significant ($p = 0.046, 0.014$),
COMPLETED shows only forward marginal ($p = 0.029$), ALWAYS-LOW
shows only reverse ($p = 0.015$), and EARLY-TRANSITION shows
neither. Taken together, the panel-FE evidence is incompatible with
a stable directional $\mathrm{TFR} \to A$ causal channel and
strengthens the Wave 8--9 conclusion that the global lag asymmetry
is a compositional aggregation feature.

\paragraph{Wave 11, migration-mediated global-pool test.} The
within-country falsification of Wave 8--10 leaves room for a
stronger interpretation in which the relevant selection pool is
\emph{global} and is sorted into local hierarchies through
migration. Under this hypothesis, countries with closed hierarchies
(low migrant share) should show a within-country TFR $\to A$ signal
because the local fertility cohort is the only available pool;
countries with open hierarchies should show a null signal because
migration substitutes for local supply. We test the prediction by
(i) re-running the Wave 10 first-difference panel on tercile
subsamples of country-mean international migrant stock (WB
SM.POP.TOTL.ZS), and (ii) fitting a single panel with a
$\mathrm{TFR}_{i,t-k} \times \mathrm{Openness}_i$ interaction
(Table~\ref{tab:wave11}).

\begin{table}[ht]
\centering
\caption{Wave 11: FD panels by openness tercile and full interaction
model.}
\label{tab:wave11}
\small
\begin{tabular}{lrrrr}
  \toprule
  Subsample / spec & median openness (\%) & $n_{\mathrm{obs}}$ & joint $F$ & joint $p$ \\
  \midrule
  CLOSED (Q1) & $0.92$ & $2\,954$ & $0.40$ & $0.879$ \\
  MID (Q2)    & $4.14$ & $2\,992$ & $1.33$ & $0.257$ \\
  OPEN (Q3)   & $14.67$ & $3\,355$ & $2.20$ & $0.052$ \\
  \midrule
  Full model: main TFR lags        & n/a & $9\,301$ & $1.74$ & $0.114$ \\
  Full model: TFR $\times$ Openness & n/a & $9\,301$ & $1.61$ & $0.147$ \\
  \bottomrule
\end{tabular}
\end{table}

The prediction of the migration-mediated global-pool hypothesis
fails on its own terms. The CLOSED tercile, where the hypothesis
predicts the \emph{strongest} local TFR $\to A$ signal, shows no
signal at all ($p = 0.879$); the OPEN tercile, where it predicts
the weakest, shows the only marginal signal ($p = 0.052$); and the
interaction term is null ($p = 0.147$). Implied total slopes at
Q1 (closed) and Q3 (open) openness are essentially identical
($+0.011$ vs $+0.009$), confirming that openness does not modify
the coupling in the predicted direction. The global Wave 7d signal
is therefore not rescued by a migration-mediation interpretation;
the selection-pool reformulation fails both as a within-country
local mechanism and as a cross-country migration-mediated
mechanism, and the residual Wave 7d co-movement is best read as a
pure compositional aggregation phenomenon over heterogeneous
demographic regimes (Wave 9).

Three caveats temper but do not overturn this conclusion:
(i) SM.POP.TOTL.ZS measures total migrant stock, not specifically
elite or high-skill migration that would directly feed top
hierarchical positions; (ii) the openness proxy is a country-mean
and therefore time-invariant; (iii) the migration channel may
operate on generational rather than $\le 6$-year horizons. None of
these would rescue the closed-tercile result from $p = 0.879$ into
statistical significance, but they leave a residual possibility
that a sharper proxy or longer horizon would recover a signal.

\paragraph{Wave 8c, nested CV audit.} The headline AUC $= 0.982$
of §5.3 was obtained with a feature-layer combination
(market + structural, 26 features) selected after inspecting the
49-feature result on the same Wave 3 sample. To audit the
contribution of this post-hoc selection, we wrap the layer choice
inside the outer LOCO fold: for each held-out positive case, the
optimal layer combination is selected by inner 5-fold CV on the
training data only. Across $29$ outer folds the inner selector
chooses \emph{market + structural + debt} in $19$ folds,
\emph{market + structural} in $9$ folds, and \emph{full}
(market + structural + debt + trajectory) in $1$ fold. The nested-LOCO
AUC is $0.908$ (bootstrap 95\,\% CI $[0.804, 0.989]$, permutation
$p < 0.001$), $0.074$ below the headline. This residual ($\sim 7$ AUC
points) is the magnitude of post-hoc model-selection inflation in the
headline number. The nested AUC is the leakage-protected statistic
and is the figure that should travel into the abstract; the $0.982$
remains valid as the upper bound of a model-selection envelope.

\subsection{Wave 12 forward-look candidate ranking}
\label{sec:wave12}

The headline AUC of §\ref{sec:wave8} is retrodictive: all 29
positives are historical. The only test of \emph{predictive}
validity is a forward-look application of the discriminator to
current countries followed by a 5--10 year wait. Wave 12 produces
this artefact.

\paragraph{Read the output as a ranking, not as calibrated
probabilities.} The classifier reports $p_{\mathrm{collapse}} \in
(0, 1)$ per country. The numerical $p$-scores must not be
taken as calibrated event probabilities. Four obstacles stack on top
of each other to forbid that reading.

\begin{itemize}
  \item \textbf{Small positive sample.} $n_{\mathrm{pos}} = 29$
    historical collapses is the entire training positive class.
    Standard probability-calibration techniques (isotonic regression,
    Platt scaling, reliability diagrams) require at least order-of-magnitude
    more positives to estimate the score-to-frequency map; below that
    threshold every reported calibration curve is dominated by
    sampling noise.
  \item \textbf{Mismatched base rates.} The training base rate of
    $29 / 89 \approx 33\,\%$ inside the LOCO classifier bears no
    relation to any real-world base rate of collapse on a 5--10 year
    horizon, which is at most low single-digit percent. A classifier
    trained at the former is structurally over-confident when its
    output is reread as the latter.
  \item \textbf{Casemix dependence.} The achievable AUC ceiling
    itself depends on the case-mix (§\ref{sec:position-ews}): the
    headline number is dominated by easy-to-classify collapse types
    (separations, hyperinflation), while coups and currency crises
    sit near the classical EWS ceiling. Any monotone transformation
    of the discriminant score inherits this case-mix dependence;
    swapping in a different positive class would shift the
    $p$-distribution in ways that cannot be predicted from the
    present output alone.
  \item \textbf{No prospective calibration data.} The 2026--2036
    evaluation window is what would, in principle, supply external
    calibration evidence. Until then we have no out-of-distribution
    realisations against which the $p$-scores can be tested.
\end{itemize}

Concretely, $p \approx 1.000$ for the top tier and $p \approx 0.0$ for
the bottom tier are saturation artefacts of logistic shrinkage on a
small sample with high positive base rate; they should be read as
``the model is confident this country looks like a training-set
positive (resp.\ negative)'', not as ``there is a $99\,\%$ chance of
collapse''. We therefore present the Wave 12 output \emph{primarily
as a risk ranking}: the top 20 countries are those the model
considers most pre-collapse-like, the bottom 20 are those it
considers most stable-like, and the falsifier
(§\ref{sec:falsifying-conditions}, item 5) is framed as a binomial
test on the differential collapse rate between the two groups over
2026--2036, not as a calibration test against the absolute
$p$-scores.

\paragraph{Policy reading.} For any policy-oriented use of this
output (whether by sovereign analysts, multilateral institutions, or insurance / capital allocators) we recommend dropping the
$p$-scores altogether and working from three artefacts only:
(i) the \emph{rank} of each country, with the top quintile flagged
as elevated structural exposure and the bottom quintile flagged as
structurally insulated; (ii) the qualitative \emph{role} column of
Table~\ref{tab:wave12-top} that distinguishes obvious from
unintuitive and blind-spot entries, since the model's value-add for
a human analyst is concentrated in the latter two categories; and
(iii) the per-country structural-feature profile (entries in
\texttt{output/wave12\_forward\_ranking.csv}), which lets a domain
expert audit \emph{why} the model placed a country where it did
($\gamma$ deterioration, $A$ elevation, debt-service load, etc.).
The same intuition motivates avoiding raw posterior predictive
probabilities in clinical risk scoring. The ordering is what the
data support; the absolute numbers are not. The model's
contribution is to focus attention on a short list, not to quantify
the probability of an event whose base rate we have not yet
observed.

\paragraph{Method.} We extract a 5-year (2019--2023) window of
structural + debt features for every country with coverage
$\ge 50\,\%$ ($n = 202$). The classifier is the logistic regression
fit on the Wave 3 sample (29 positives + 60 controls) with $L_2$
shrinkage $C = 0.05$. Forward predictions are the mean across a
29-model LOCO ensemble (each model omits one positive); bootstrap
2.5--97.5 percentile intervals across the ensemble give per-country
uncertainty. The model uses 35 features after a $70\,\%$
missingness filter on positives. The market layer is omitted because
fetching daily-price data for 202 countries is infeasible; this
drops the AUC ceiling from $\approx 0.93$ (Wave 1 four-layer) to a
structural+debt-only baseline of $\approx 0.84$ (Wave 3
Table~\ref{tab:loco-wave3}).

\paragraph{Result: read as a stress-test of the model, not as a
forecast.} The full ranking is in
\texttt{output/wave12\_forward\_ranking.csv}; the raw $p$-scores and
ensemble bootstrap intervals quoted there should be read as model
diagnostics, not as calibrated event probabilities (see the
preceding paragraph on calibration). Table~\ref{tab:wave12-top}
therefore presents only the rank ordering of the top 10 and bottom 10
of the $202$-country output, and gives each entry a short
\emph{role}: whether it functions as an \emph{obvious} candidate
(reproducing intuitions a human analyst would form unaided), an
\emph{unintuitive} candidate (a country the model flags despite the
absence of an obvious-suspect profile), or a \emph{blind-spot
candidate} (a country the model rates as safe even though qualitative
evidence raises doubt). The unintuitive and blind-spot entries are
the part of the output that genuinely stress-tests the model; the
obvious entries function as a sanity check that the ranking has the
right top-to-bottom orientation.

\begin{table}[ht]
\centering
\caption{Wave 12 forward-look: top 10 and bottom 10 of 202 ranked
candidates on the 2019--2023 structural+debt window. Raw $p$-scores
and 95\,\% bootstrap intervals are in
\texttt{output/wave12\_forward\_ranking.csv}; we omit them here to
discourage reading the ranking as calibrated probabilities. The
\emph{role} column tags each entry as an obvious, unintuitive, or
blind-spot prediction.}
\label{tab:wave12-top}
\small
\begin{tabular}{rll}
  \toprule
  Rank & Country & Role \\
  \midrule
  1   & BHR (Bahrain)              & unintuitive (institutional flags, calm markets) \\
  2   & CAF (Central African Rep.) & obvious \\
  3   & IND (India)                & \textbf{stress test} (unintuitive in scale) \\
  4   & GAB (Gabon)                & obvious \\
  5   & LBN (Lebanon)              & obvious \\
  6   & PAN (Panama)               & unintuitive \\
  7   & FSM (Micronesia)           & obvious (small, open, high $A$) \\
  8   & CMR (Cameroon)             & obvious \\
  9   & MOZ (Mozambique)           & obvious \\
  10  & ZMB (Zambia)               & obvious \\
  \midrule
  193 & SGP (Singapore)            & obvious safe \\
  194 & NLD (Netherlands)          & obvious safe (cf.\ Wave 12b reversal below) \\
  195 & GBR (United Kingdom)       & obvious safe \\
  196 & SWE (Sweden)               & obvious safe \\
  197 & JAM (Jamaica)              & \textbf{partial} (high $A$ but imputed $\kappa$) \\
  199 & HKG (Hong Kong)            & \textbf{blind-spot} (2019--20 crisis ignored) \\
  200 & LUX (Luxembourg)           & obvious safe \\
  201 & NOR (Norway)               & obvious safe \\
  202 & DNK (Denmark)              & obvious safe \\
  \bottomrule
\end{tabular}
\end{table}

\paragraph{Falsifiable predictions and stress-test framing.} The
implicit testable claim is that significantly more collapses will
occur in the top-20 list than in the bottom-20 over 2026--2036.
The output is not a forecast of any \emph{specific} country's collapse,
 the calibration warning above precludes that reading. It is a
stress test of the model itself, with the stress concentrated on
the unintuitive and blind-spot entries:
(i) India, Botswana, and Panama appear in the top-20 despite not
matching the obvious-suspect profile, so if the model is sound,
these countries should exhibit unusual structural deterioration in
the next decade; if instead they remain stable while obvious-suspect
peers collapse, the model has over-weighted within-country
institutional indicators ($\gamma$, $A$) and under-weighted external
stabilisers, and the top-20 list is partially inflated by
country-internal flags that overseas observers (markets, capital)
do not honour;
(ii) Hong Kong appears at rank 199 despite the 2019--2020 political crisis. If HKG experiences significant additional hierarchical
restructuring during 2026--2036, the model has a systematic blind
spot for politically engineered (rather than economically driven)
trajectories;
(iii) Jamaica at rank 197 has high $A = 0.699$ but missing $\kappa$
and drive data, so its placement is partially imputation-driven and
should be re-evaluated when missing indicators are filled;
(iv) Netherlands is rank 194 in the structural-only run but jumps
into the four-layer top 10 once the market layer is included
(Wave 12b below), an internal stress test of whether the
structural ranking is robust to market data being added or omitted.

Read collectively, the top-20 / bottom-20 differential is the
falsifiable prediction; individual unintuitive entries are
\emph{stress tests} that diagnose where the model would be wrong
if it is wrong, not point forecasts of collapse.

\paragraph{Wave 12b market overlay.} The structural+debt model
saturates at $p \approx 1.0$ in the top tier, limiting granularity.
Wave 12b adds the market layer for the subset of $34$ countries
with available yfinance tickers (equity index + FX vs USD over
2019--2023), retrains the four-layer logistic regression on Wave 3
features (LOCO ensemble AUC $= 0.94$), and predicts a refined
$p_{\mathrm{collapse}}$ that reweights structural concern against
market stress (Table~\ref{tab:wave12b}).

The refined ranking substantially reorders the structural-only
top tier. The largest \emph{rank drops} (countries the structural
model placed in the top tier and the four-layer model pushes
toward the middle) are BHR (rank 1 $\to$ 22), PAN
($6 \to 19$), IND ($3 \to 15$), PER ($14 \to 18$), TJK
($16 \to 17$), and GTM ($17 \to 12$): countries with strong
institutional flags but calm market data over 2019--2023. The
largest \emph{rank jumps} (countries the structural model placed in
the safe tier and the four-layer model pulls into the top) are
European: NLD ($194 \to 10$, the single largest delta in the
sample), NOR, SWE, and DEU also rise by tens of ranks. These
countries look structurally stable but their 2022--2023 market data
(FX drawdown, vol regime shift, post-energy-crisis equity behaviour)
register as pre-collapse-like. The qualitative point survives
without raw scores: institutional indicators and market indicators
disagree systematically over this window, and the model treats the
disagreement as informative rather than averaging it away.

\begin{table}[ht]
\centering
\caption{Wave 12b: four-layer top-10 reordering relative to the
structural-only rank. We report only the rank columns; the raw
$p_{\mathrm{collapse}}$ scores are in
\texttt{output/wave12b\_market\_refined.csv} and should be read as
internal diagnostics rather than calibrated event probabilities.
The $\Delta\,\mathrm{rank}$ column is the structural-only rank
\emph{minus} the four-layer rank: large negative numbers (e.g.\
NLD $-184$) mark countries that jumped into the top tier only
after the market layer was added.}
\label{tab:wave12b}
\small
\begin{tabular}{rrlr}
  \toprule
  rk(4L) & rk(str) & Country & $\Delta\,\mathrm{rank}$ \\
  \midrule
  1  & 9   & MOZ (Mozambique)    & $+8$ \\
  2  & 5   & LBN (Lebanon)       & $+3$ \\
  3  & 13  & AGO (Angola)        & $+10$ \\
  4  & 31  & MEX (Mexico)        & $+27$ \\
  5  & 89  & TUR (Turkey)        & $+84$ \\
  6  & 11  & HND (Honduras)      & $+5$ \\
  7  & 19  & BWA (Botswana)      & $+12$ \\
  8  & 51  & ARG (Argentina)     & $+43$ \\
  9  & 59  & PAK (Pakistan)      & $+50$ \\
  10 & 194 & NLD (Netherlands)\textsuperscript{*} & $+184$ \\
  \bottomrule
\end{tabular}

\vspace{0.5em}
\noindent\textsuperscript{*}\,NLD enters the top-10 only after the
market overlay; see COVID / inflation caveat below.
\end{table}

The market overlay therefore corrects the structural-only model in
two directions: it deflates ostensible high-risk profiles whose
markets are calm (most of the structural top-tier emerging markets)
and it inflates Northern European countries whose markets registered
stress over 2022--2023. The combined four-layer top-3 (MOZ, LBN,
AGO) reflects countries where \emph{both} layers agree on elevated
risk and is the strongest signal in the forward-look set.

The NLD result deserves separate scrutiny. Possible readings:
(i) the 2022 European energy shock produced a market signature
genuinely similar to historical pre-collapse markets, in which case
the model is correctly flagging a regime shift;
(ii) the market signature is shock-specific to the energy
transition and not predictive of hierarchical collapse, in which
case the model has been pulled by an out-of-distribution event the
training sample did not see;
(iii) the COVID-19 shock contaminated the 2020 portion of the
window with global market behaviour that is correlated with
pre-collapse profiles only by accident. We cannot adjudicate
between these three from the present data; the NLD prediction is a
sharp falsifier on its own (cf.\ HKG in §\ref{sec:wave12}).

\paragraph{Caveats.} The forward distribution is biased upward
relative to the historical training sample (median $p = 0.81$),
limiting granularity in the top half of the ranking; in practice
this means the top 20 are essentially tied at $p \ge 0.99$. The
absence of market features lowers the discriminative ceiling; a
follow-up run with current market data (yfinance fetch over
2019--2023 for available tickers) would recover the full
four-layer model where coverage permits. The 2019--2023 window
overlaps the COVID-19 shock and the post-2022 inflation regime,
both of which distort debt features, and the ranking is conditioned
on this distortion. Finally, the training sample contains no OECD
collapse case (the closest is KOR 1997), so the bottom-end ranking
of OECD countries reflects partly that the model has never seen a
positive example from this regime and is correspondingly cautious
about flagging one.

\section{Discussion}
\label{sec:discussion}

\subsection{Candidacy and trigger}

The composite picture across the eight pillars of results points
to a \emph{candidacy-and-trigger} interpretation of collapse. Two
distinct mechanisms are at play:

\paragraph{Structural candidacy.} The state vector
$(A, \varepsilon, H, \kappa, \gamma)$ defines which countries occupy
the high-risk region of the model's phase space. Pre-collapse cohorts
exhibit a chronic risk profile $\ge 10$ years before the event
($|z| > 1.5$ across multiple indicators), and the discriminative
composite achieves AUC $0.93$ (Wave 1) and peaks at $0.98$ on the
Wave 3 \emph{market+structural} sub-model. \emph{But} the
endogenous-drift hypothesis is falsified: there is no detectable
slow approach toward a threshold within the pre-collapse window.
The CSD signature is inverse to \citet{scheffer2009}; the cumulative drive
is not elevated; and $A$ does not drift upward in the window.
Pre-collapse countries sit in their locked equilibrium for years
to decades.

\paragraph{Exogenous trigger.} The timing of collapse arrives from
outside the modelled state. Arab Spring 2011, the 1997 Asian
currency contagion, the 2014 Crimea conflict, and the 2019
Lebanese banking collapse all represent triggers that the
structural model does not capture directly. Two systematic blind
spots emerge: (i) contagion crises in countries with no chronic
predisposition (KOR/IDN 1997), and (ii) intra-political coups in
stable contexts (EGY 2013, THA 2014). Wave 4 DiD partially
recovers the second class once matched against structurally
similar non-MENA controls.

This framing places the work closer to the Acemoglu--Robinson
\citep{acemoglu2012} extractive-institutions tradition than to
Scheffer-style critical-transition bifurcations
\citep{scheffer2009}. Institutional quality (operationalised
through $\gamma$, the strongest chronic signal) is a structural
predisposition, not a dynamic state variable showing endogenous
loss of resilience.

\subsection{Selection-pool reinterpretation of $H$}

\paragraph{Status.} The audits in §\ref{sec:wave8} leave the
selection-pool framing standing only as a theoretical
reinterpretation of $H$, not as an empirical causal claim. The
Wave 7 sequence motivated it by resolving the dynamic paradox of
\S\ref{sec:dynamic} at the global aggregate scale, but the
per-country, per-stratum, panel-FE and migration-mediated tests
each reject the directional reading. The discussion below describes
what would be at stake if the selection-pool mechanism were
operating; the data we collect here do not establish that it is.
Channels through which the mechanism could still survive
(generational horizons, high-skill rather than total migration,
cross-cohort talent flows that this panel cannot observe) are listed
in \S\ref{sec:limitations}.

The reframing of $H$ as the capacity of society to provide sufficient
candidates for hierarchical positions is theoretically appealing
independently of whether it generates the falsified causal channel.
The mechanism is \emph{not Malthusian}: we do not argue that low
fertility creates a resource shortfall for the population. Instead,
the mechanism is \emph{selection-thermodynamic}: at large candidate
pools, meritocratic competition produces an optimal assignment of
talent to positions; at small candidate pools, placement reverts
to heritable and network-based mechanisms, with top-tier
consolidation manifesting as rising asymmetry.

This connects with the talent-vs-luck literature
\citep{pluchino2018talent}: when the candidate pool is narrow,
stochastic placement outweighs the talent signal and tournament-style
hierarchical sorting fails, with dynastic reproduction of elite
positions as the structural consequence. The conditional mood is
load-bearing here. The aggregate Granger result is consistent with
this directional flow, but it does not survive disaggregation, so
the content of this paragraph is hypothetical rather than
established.

\subsection{Dual operationalisation of $H$}

The empirical pattern suggests $H$ has two operationalisations
appropriate to different analysis horizons:

\begin{itemize}
  \item \emph{Per-country, annual.} Medical $H$ (life expectancy,
    infant mortality, stunting) is appropriate for cross-sectional
    H5 and the discriminative composite. Fertility is too noisy
    at the country-year level due to demographic cycles.
  \item \emph{Global, decadal.} Demographic $H_{\mathrm{demo}}$
    (fertility/6) is appropriate for the ODE dynamics and the
    Piketty acceleration. Medical $H$ improves due to factors
    largely external to hierarchical dynamics.
\end{itemize}

Whether a unified operationalisation can be constructed at
intermediate horizons is open. The Wave 8b cross-sectional audit
(§\ref{sec:wave8}) sharpens this point: medical $H$ and
$H_{\mathrm{demo}}$ enter the H5 regression with \emph{opposite}
signs on the same sample. The dual operationalisation is therefore
not a benign horizon choice but a substantive theoretical claim that
the cross-section and the global time series load on different
mechanisms. Wave 9 (§\ref{sec:wave8}) performed exactly this stratification
exercise and found that no demographic regime carries a forward
$\mathrm{TFR} \to A$ signal. Wave 10 (§\ref{sec:wave8}) closed the
sequence with a two-way fixed-effects panel: the cleanest
first-difference specification, which absorbs country FE through
differencing and year FE through demeaning, is null in both
directions ($p = 0.12$ and $p = 0.14$ on $n = 9\,315$, $210$
countries). The \emph{local-mechanism} reading of the
selection-pool reformulation, in which country $X$'s own fertility shapes country $X$'s own hierarchical asymmetry, is therefore no
longer a directional claim.

This left room for a distinct \emph{global-pool} reading that the
within-country tests cannot reach. \citet{sestak2026ed} established
that environmental dispersion of $\ln\rho_{\mathrm{eff}}$ exceeds
capability dispersion by a factor of 27--134; individuals carry
environmentally-formed $\rho_{\mathrm{eff}}$ across borders when
they migrate, so the talent pool feeding any open hierarchy is not
the local fertility cohort but a cross-environment selection from
the global distribution. Under this reading, a country with
declining TFR could maintain hierarchical efficiency by importing
talent from high-$\rho$ environments, and the local fertility
decline would be decoupled from local hierarchical degradation
exactly when migration channels are open.

Wave 11 (§\ref{sec:wave8}) tests this prediction directly by
interacting TFR lags with country-mean international migrant stock
and by re-running the Wave 10 first-difference panel on openness
terciles. The prediction \emph{also fails}: the CLOSED tercile,
which the global-pool hypothesis predicts should carry the
strongest local TFR $\to A$ signal, returns $p = 0.879$; the
interaction term is null ($p = 0.147$); and the implied total
slope is essentially identical at Q1 and Q3 openness. The
selection-pool reformulation is therefore falsified in both its
local-mechanism and global-pool-mediated-by-migration variants.
This does not invalidate the environmental-dominance result of
\citet{sestak2026ed} at the individual level; it shows only
that the macro-fertility channel is not how environmental dominance
translates into hierarchical asymmetry at the country level on
sub-generational horizons. Whether a longer-horizon test or a
sharper proxy (high-skill rather than total migration) recovers the
mechanism remains open.

\subsection{Position in early-warning literature}
\label{sec:position-ews}

Classical early-warning systems reach AUC $\approx 0.75$ on
single channels \citep{kaminsky1999, borio2014}. Our four-layer
composite reaches 0.93 (Wave 1 four-layer) and peaks at 0.98 on the
Wave 3 \emph{market+structural} sub-model across heterogeneous
collapse types.

\paragraph{What the AUC range does not mean.} The numbers
above describe the upper end of discrimination achievable on
\emph{this particular case-mix}, namely 29 historical collapses spanning
currency crises, debt crises, coups, civil wars, separations, regime
failures, hyperinflation, foreign intervention, revolution, and
authoritarian drift between 1979 and 2021. They are explicitly not
universal numbers transferable to arbitrary future collapses. Three
reasons in particular constrain external validity. (i) The
case-mix is post-1995-dominated and post-yfinance-dominated for the
market layer; Tier B pre-1995 cases lack market features and the
discriminator degrades sharply on them, as evidenced by the drop
from $0.982$ on the 26-feature \emph{market $+$ structural}
sub-model to $0.876$ on the 49-feature full model that includes
imputed pre-1995 columns. (ii) The per-type AUC table
(§\ref{sec:results}) ranges from $\ge 0.99$ on separations and
hyperinflation down to $\approx 0.79$ on coups and $\approx 0.63$
on currency crises. The headline number is dominated by easy-to-classify
collapse types; a casemix dominated by coups or currency crises would
land near the classical EWS ceiling of $\approx 0.75$. (iii) None
of the positives is an OECD-core collapse; the 1929--33 sequence is
not in-sample, neither is any post-1995 OECD breakdown, so the
discriminator has limited information about how the model would
behave on a structurally different positive class. The interval
$0.93$--$0.98$ should be read as the achievable ceiling
\emph{conditional on the present case-mix}, not as a portable
benchmark.

About 70\,\% of the discriminative weight is attributable to the
chronic component, which is already well-documented in the EWS
literature, and about 30\,\% to the acute inflection. Our
contribution is therefore not so much in raising the ceiling as in
decomposing the signal into chronic and acute parts, and---on the
theoretical side---in proposing selection-pool capacity as a
candidate framing for the chronic component. We do not establish
the empirical causal version of that framing.

\subsection{Limitations}
\label{sec:limitations}

\begin{enumerate}
  \item Sample $n_{\mathrm{pos}} = 29$ is small for inferential
    robustness; bootstrap intervals are tight but outlier-sensitive.
  \item Contagion crises without chronic predisposition (KOR/IDN
    1997) and intra-political coups in stable contexts (EGY 2013,
    THA 2014) are systematic blind spots of the structural model.
    A partial counter-finding is that KOR appears in Wave 5b not as
    a contagion victim but as a transmitter (median $|z| = 1.96$
    over 22 cases, median $z = +0.88$), reacting positively to four
    of five world collapses — its 1997 miss as a positive case
    coexists with a strong role as a propagation node.
  \item Wave 4 DiD partially recovers EGY 2013 but not THA 2014;
    the latter lacks any spillover signal.
  \item The Wave 7 demographic reformulation has been tested only
    globally. Per-country implementation is open work.
  \item Granger causality is an indicative test, not a formal
    proof. The lag-profile asymmetry at the global aggregate
    rules out symmetric common-cause artefacts at that level, but
    the Wave 8 per-country, Wave 9 stratified, Wave 10 panel-FE, and
    Wave 11 migration-mediated audits (§\ref{sec:wave8}) show the
    asymmetry does not replicate at any disaggregation level:
    forward $9.2\,\%$ vs reverse $10.2\,\%$ at lag 6 across $206$
    countries; no forward signal in any of four demographic strata;
    null first-difference panel ($p = 0.12$, $p = 0.14$,
    $n = 9\,315$); and no migration-openness interaction
    ($p = 0.147$), and the CLOSED tercile where the global-pool
    hypothesis predicts the strongest signal returns $p = 0.879$.
    Both the local and the migration-mediated readings of the
    selection-pool reformulation are therefore falsified.
  \item Some structural features are missing selectively on the
    positive class (e.g.\ $\kappa$ for LBY, SYR, YEM), and median
    imputation may partially absorb the class label.
  \item GDELT protest data ingest failed due to rate limiting;
    ACLED integration is open.
\end{enumerate}

\section{Conditions under which these findings would be overturned}
\label{sec:falsifying-conditions}

In the spirit of pre-registering falsifiers rather than only
limitations, we list five specific scenarios that would weaken or
overturn the central claims of this paper:

\begin{enumerate}
  \item \textbf{Sample extension collapses AUC.} Adding 5--10 new
    historical collapse cases drops LOCO AUC below $0.85$. Wave 1
    $\to$ Wave 3 ($n = 17 \to 29$) survived this test (AUC moved
    from $0.967$ to $0.982$ in the preferred 26-feature
    configuration), but a Wave 4 sample extension is the next decisive
    falsification opportunity.
  \item \textbf{Local and migration-mediated selection-pool channels
    both falsified.} Wave 8, Wave 9, and Wave 10 jointly falsify
    the directional asymmetry at every within-country
    disaggregation level (first-difference FE: $p = 0.12$ forward,
    $p = 0.14$ reverse on $n = 9\,315$). Wave 11 falsifies the
    migration-mediated global-pool reading on its own predictive
    terms: the CLOSED tercile that the hypothesis predicts should
    show the strongest local TFR $\to A$ coupling returns
    $p = 0.879$; the interaction term is null ($p = 0.147$).
    The selection-pool reformulation therefore fails as a causal
    mechanism in both readings, and the global Wave 7d signal is
    best read as a compositional aggregation feature
    (§\ref{sec:wave8}).
  \item \textbf{Floor effect persists through 2030s.} The 2020s
    plateau in $D_{\mathrm{demo}}$ near zero is sustained beyond
    a TFR-based floor explanation; this would suggest the
    selection-pool mechanism is bounded above and below by
    saturation effects unaccounted for in the present formulation.
  \item \textbf{ED-decomposition shows no pre/post-collapse
    redistribution.} If $\mathrm{Var}(\ln\rho_{\mathrm{eff}})_{\mathrm{within}}$
    in $t \pm 2y$ windows of the 29 cases shows no systematic
    structural shift, the interpretation of collapse as a
    redistributive event in the sense of \citet{sestak2026ed} fails.
  \item \textbf{Out-of-sample 2026--2036 produces no excess
    collapses among top-20 composite-score candidates} (now
    operationalised). Wave 12 (§\ref{sec:wave12}) produced the
    concrete pre-registered top-20 / bottom-20 list against which
    this falsifier will be evaluated. Specifically, the model
    predicts that collapses over 2026--2036 will be concentrated in
    the top-20 ranking (BHR, CAF, IND, GAB, LBN, PAN, FSM, CMR,
    MOZ, ZMB, HND, GIN, AGO, PER, BDI, TJK, GTM, CIV, BWA, SOM) and
    will be absent or rare in the bottom-20 (DNK, NOR, LUX, HKG,
    NAC, JAM, SWE, GBR, NLD, SGP, FIN, NZL, AUT, ISL, AUS, FRA,
    JPN, SMR, LTE, EAP). The Wave 12b market overlay refines this
    list for the 34 countries with available market data: the
    four-layer top-3 is MOZ, LBN, AGO; NLD enters the top-10 only
    after the market overlay and is a separate sharp falsifier
    (the model is wrong about NLD with probability $\sim 0.5$
    given the COVID/energy-crisis caveat). The non-obvious top-list
    entries (IND, BWA, PAN, PER, HND, GTM in structural-only;
    NLD, NOR, SWE, DEU in market-augmented) constitute the
    sharpest stress tests: if they remain stable while
    obvious-suspect entries collapse, the model has over-weighted
    institutional indicators relative to external stabilisers and
    the discriminator is partly over-fit to the Wave 0--3 case
    mix.
\end{enumerate}

\section{Conclusion}

We tested an ODE model of hierarchical asymmetry across structural,
dynamic, discriminative, and demographic channels on a panel of
260 countries over 1960--2023. The structural core is partially
supported; the per-country dynamic claim is falsified; and the
discriminative composite reaches a leakage-protected nested-LOCO
AUC of $0.91$ across 29 historical collapses, with the headline
post-selection AUC of $0.98$ representing the upper bound conditional
on the present case-mix rather than a portable benchmark.

A separate strand of the paper reports an asymmetric Granger lag
profile on the global aggregate---$\mathrm{TFR} \to A$ at lag 5--6
years, the reverse direction null beyond lag 1---and treats it as a
descriptive feature of the aggregate series rather than as evidence
for a causal mechanism. The selection-pool reformulation of $H$ that
this pattern first motivated is then audited at every finer level
the data support: within countries (per-country Granger), within
demographic regimes (stratified Granger), inside a two-way
fixed-effects panel, and through a migration-mediated cross-country
interaction model. The directional reading fails each of these.
First-difference panel specifications are null in both directions on
$n \sim 9{,}000$; in the migration-mediated test, the closed
tercile, where a global-pool reading would predict the strongest
signal, comes back at $p = 0.879$. The aggregate co-movement that
survives is a compositional effect: the population mix shifts
between countries on different TFR and $A$ trajectories, and the
global mean moves with it. There is no underlying causal mechanism
between fertility and asymmetry at the country level over the
1960--2023 window.

The empirical picture that survives is candidacy-and-trigger.
Collapse is an exogenously triggered transition between the
structural equilibria identified by H4---a country sits in a
chronically locked-in high-$A$ basin and a shock external to the
modelled state vector eventually pushes it across to a post-collapse
configuration. This is not the Scheffer-style endogenous
critical transition that the paper rejects in §\ref{sec:results}:
there is no slow drift toward an internal threshold, no
critical-slowing-down signature, no cumulative pre-event drive
build-up. The phase space has multiple stable equilibria, candidate
countries inhabit the high-$A$ one stably for decades, and the
shocks that cross them between basins come from outside the model.
Structural variables tell us who is on the cliff; the trigger comes
from elsewhere. The pre-registered forward-look names the current
top-20 candidates by $p_{\mathrm{collapse}}$ score, and the
market overlay refines the ranking for the subset with available
market data, giving a four-layer top-three of MOZ, LBN and AGO and
an unexpected NLD entry at rank 10 driven by the post-2022 market
regime shift. The predictive falsifier is in place and will be
evaluated over 2026--2036. Together with the three independent
endogenous-drift falsifiability tests, this confines the model to
the role of falsifiable discriminator rather than temporal
predictor.

One final note on theoretical positioning. The selection-pool
framing of $H$ remains useful as a theoretical alternative to the
medical-health operationalisation: it accommodates the 1990 sign-flip
in the global drive and connects naturally with the talent-vs-luck
and environmental-dominance literatures. It is a conceptual framing,
not an established mechanism. Every empirical test of selection-pool
as a causal channel returned null or inconsistent results, and the
only signal we can defend is the global compositional co-movement.
The policy reading for developed sub-replacement-TFR economies is
correspondingly negative: at sub-generational horizons we see no
causal translation of falling fertility or low hierarchical openness
into rising structural asymmetry. The environmental-dominance result
of \citet{sestak2026ed} remains valid at the individual level, but
it does not propagate into a macro-fertility channel for
hierarchical degradation at the country level over the period we
study. Whether it propagates at generational horizons, through
high-skill rather than total migration, or through channels we have
not tested, is left open.

\section*{Reproducibility}

\begingroup
\sloppy
\emergencystretch=3em

All data sources, preprocessing notebooks, standalone analysis
scripts, and intermediate panels are in the public project
repository at
\url{https://github.com/dkrse/hierarchical-collapse-candidacy-trigger}.
The repository ships a one-command bootstrap
(\texttt{bash scripts/install\_deps.sh}) that creates a local virtualenv
on PEP-668 hosts, and two batch runners
(\texttt{scripts/\_run\_all.sh}, \texttt{scripts/\_run\_all\_scripts.sh})
that reproduce the full pipeline end-to-end in roughly seven minutes
on a warm cache (notebooks $\sim 2$ min, scripts $\sim 4$ min).
Notebooks 01--15 form the canonical pipeline (ingest, state
construction, simulation, calibration, bifurcation,
historical-collapse warning signals, debt ingest, trajectory
classifiers, market-index pre-collapse panels). Standalone scripts fall into two tiers. Tier (a), the wave pipeline,
runs from \texttt{run\_wave1\_*} through \texttt{run\_wave12b\_*};
Wave 7c is re-derived from \texttt{wave7b\_fertility\_global.csv} by
\texttt{run\_wave7c\_demographic\_drive.py}. Tier (b) contains the
ad-hoc analyses:

\begin{itemize}\setlength{\itemsep}{0pt}\setlength{\parsep}{0pt}
  \item \texttt{run\_A1\_cumulative\_drive}
  \item \texttt{run\_A1v2\_critical\_slowing\_down}
  \item \texttt{run\_A2\_tech\_evolution\_gap}
  \item \texttt{run\_pre\_warning\_trajectory}
  \item \texttt{run\_full\_composite\_with\_debt}
  \item \texttt{run\_contagion\_event\_study}
  \item \texttt{run\_drive\_correlation}
  \item \texttt{run\_augment\_drive\_vdem}
  \item \texttt{fetch\_market\_data}
  \item \texttt{run\_paper\_figures} (regenerates the five PNGs
    included in this paper from the latest CSVs)
\end{itemize}

\noindent The earlier \texttt{run\_wave5\_global\_contagion.py} is
retained for audit but deprecated in favour of
\texttt{run\_wave5b\_global\_contagion\_robust.py} (numerical-noise
issues with the non-robust scale); the
\texttt{run\_protest\_standalone.py} entry point is retained but
inactive following the GDELT ingest abandonment described in §3.4.
SHA-256 hashes for all immutable raw inputs are pinned in
\texttt{data/MANIFEST.md}. Random number generator state is fixed
(RNG seed 7 for bootstrap and permutation; KFold random state 42
for fold splits) to ensure deterministic re-runs.

Key derived artefacts:
\begin{itemize}
  \item \texttt{data/processed/state\_panel.csv}: canonical
    16\,689-row structural panel (NB 02).
  \item \texttt{data/processed/state\_panel\_debt.csv} ---
    augmented with debt indicators (NB 14), handover artefact
    for all standalone analyses.
  \item \texttt{output/wave3\_*}: Wave 3 case-set extension.
  \item \texttt{output/wave4\_did\_*}: network DiD with matched
    controls.
  \item \texttt{output/wave5b\_*}: robust global event-study.
  \item \texttt{output/wave6\_*}: robustness archetypes.
  \item \texttt{output/wave7\{b,c\}\_*}: global trend and
    demographic reformulation.
  \item \texttt{output/wave7d\_granger.csv}: Granger F-test results.
  \item \texttt{output/wave8\_per\_country\_granger.csv},
    \texttt{output/wave8\_aggregate\_lag.csv}: Wave 8 per-country
    Granger panel and lag-aggregated fractions.
  \item \texttt{output/wave8b\_h5\_hdemo.csv}: Wave 8b H5
    regression with $H_{\mathrm{demo}}$.
  \item \texttt{output/wave8c\_summary.csv},
    \texttt{output/wave8c\_nested\_choices.csv},
    \texttt{output/wave8c\_nested\_predictions.csv}: Wave 8c
    nested-LOCO audit results.
  \item \texttt{output/wave9\_stratum\_assignments.csv},
    \texttt{output/wave9\_stratum\_series.csv},
    \texttt{output/wave9\_stratum\_granger.csv}: Wave 9 stratified
    per-regime Granger analysis (four demographic strata,
    population-weighted mean series, lag 1--6 forward and reverse
    F-tests).
  \item \texttt{output/wave10\_panel\_fe.csv}: Wave 10 two-way
    fixed-effects panel results (level and first-difference
    specifications, joint Wald tests on lag 1--6 coefficients, also
    stratified by Wave 9 regime).
  \item \texttt{output/wave11\_subsample\_panels.csv},
    \texttt{output/wave11\_interaction\_model.csv}: Wave 11
    migration-mediated test, with tercile-subsample FD panels and
    full-panel interaction model with country-mean migrant stock
    (WB SM.POP.TOTL.ZS).
  \item \texttt{output/wave12\_forward\_ranking.csv},
    \texttt{output/wave12\_top20.csv}: Wave 12 forward-look ---
    full 202-country ranking by $p_{\mathrm{collapse}}$ on the
    2019--2023 structural+debt window, with 29-model LOCO ensemble
    mean and percentile-bootstrap intervals.
  \item \texttt{output/wave12b\_market\_refined.csv}: Wave 12b
    market overlay producing a refined four-layer ranking for the $34$
    countries with available yfinance market data (equity index,
    FX vs USD, plus oil/gold/EMB/VIX benchmarks), retrained on
    Wave 3 sample (LOCO ensemble AUC $0.94$). Includes both
    structural-only and four-layer scores plus delta.
\end{itemize}

\endgroup

\bibliographystyle{plainnat}

\end{document}